\documentclass[prl,showpacs,twocolumn,amsmath,amssymb,floatfix,superscriptaddress]{revtex4-2}
\usepackage{graphicx} 
\usepackage{bm}
\usepackage{float}
\usepackage{bbm}
\usepackage{color}
\usepackage{amsmath}
\usepackage{amssymb} 
\usepackage{color}
\usepackage{footnote}
\usepackage{comment}
\usepackage{hyperref}
\usepackage{subfigure}
\usepackage{appendix}
\usepackage[applemac]{inputenc}
\usepackage[T1]{fontenc}

\begin{document}
\title{Directional Andreev-Reflection Signatures of Inter-Orbital Pairing in Sr$_2$RuO$_4$}

\author{G. Csire*}
\affiliation{CNR-SPIN, c/o Universit\`a di Salerno, IT-84084 Fisciano (SA), Italy}

\author{Y. Fukaya*}
\affiliation{Faculty of Environmental Life, Natural Science and Technology, Okayama University, 700-8530 Okayama, Japan}

\author{M. Cuoco}
\affiliation{CNR-SPIN, c/o Universit\`a di Salerno, IT-84084 Fisciano (SA), Italy}

\author{Y. Tanaka}
\affiliation{Department of Applied Physics, Nagoya University, Nagoya 464-8603, Japan}

\author{R.K. Kremer}
\affiliation{Max-Planck Institute for Solid State Research, Stuttgart, Germany}

\author{A.S. Gibbs}
\affiliation{University of St Andrews, St Andrews, Scotland, UK}

\author{G.A. Ummarino}
\affiliation{Department of Applied Science and Technology, Politecnico di Torino, Torino, Italy}

\author{D. Daghero}
\affiliation{Department of Applied Science and Technology, Politecnico di Torino, Torino, Italy}

\author{R.S. Gonnelli*}
\affiliation{Department of Applied Science and Technology, Politecnico di Torino, Torino, Italy}


\begin{abstract}
Unconventional superconductivity in quasi--two--dimensional systems is commonly identified through the emergence of Andreev bound states (ABS) at in--plane edges, while surfaces perpendicular to out--of--plane direction remain fully gapped due to weak interlayer coherence. This directional anisotropy has long served as a key paradigm for constraining pairing symmetries. Here, we show that Sr$_2$RuO$_4$ exhibits a striking reversal of this behavior. Using edge- and surface-sensitive spectroscopy, we observe pronounced in-gap ABS at surfaces perpendicular to the out-of-plane direction, whereas in-plane edges exhibit a reduced intensity of the in-gap spectral features.
We show that this anomalous anisotropy can arise from the inter-orbital character of the superconducting pairing. Both even- and odd-parity inter-orbital pairing channels naturally generate robust surface ABS while suppressing planar edge modes and can also provide a mechanism for the appearance of a horizontal line node.  
Supported by \textit{ab initio} and model calculations, including Sr$_2$RuO$_4$/Ag interface reconstructions, our results highlight the possible role of inter-orbital correlations in shaping the spectroscopic response and provide constraints on the structure of the superconducting order parameter in Sr$_2$RuO$_4$.
\end{abstract}

\maketitle


\textit{Introduction.--} Unconventional superconductivity~\cite{UedaSigristReview,Tsuei2000} arises from nonstandard pairing symmetries and complex order parameters. These phases can host anisotropic responses~\cite{Kashiwaya_2000,Hu1994,TK95} and topological states~\cite{Kitaev_2001,Qi2011,Sato_2017,Tanaka2012,Tanaka2024}, with potential applications in quantum computing and quantum electronics~\cite{Kitaev_2001,Ivanov2001,Sato2003,Nayak2008,Sau2010,Alicea2011}. 
A defining feature is the emergence of Andreev bound states (ABSs) at boundaries, which have been observed in a wide range of materials~\cite{Kashiwaya1995,Alff1997,   Kashiwaya_2000,LESUEUR1992325,CovingtonPRL1997,WeiPRL1008,BiswasPRL2002,Kashiwaya2011,RourkePRL2005,DagheroNatCommun2012,WaltiPRL2000,Liu2019,Zhu2023,Yano2023,Yoonnpj2024,Wang2025,bisset2025}. These states originate from sign-changing superconducting order parameters and are protected by topology~\cite{Kitaev_2001,Qi2011,Sato_2017,Tanaka2012}.

Layered quasi-two-dimensional (quasi-2D) superconductors provide a natural platform to study these effects. Because pairing and quasiparticle motion are largely confined to the planes, strong anisotropy emerges. Here, in--gap ABSs are expected mainly at in-plane edges where superconducting coherence is interrupted, while weak interlayer coupling suppresses bound states at out-of-plane surfaces. Consequently, unconventional pairing is typically expected to manifest itself as enhanced in-gap conductance in in-plane spectra, whereas out-of-plane transport remains gapped.

\begin{figure*}[!t]
   \centering
    \vspace{-2cm}
    \includegraphics[width=1.00\linewidth]{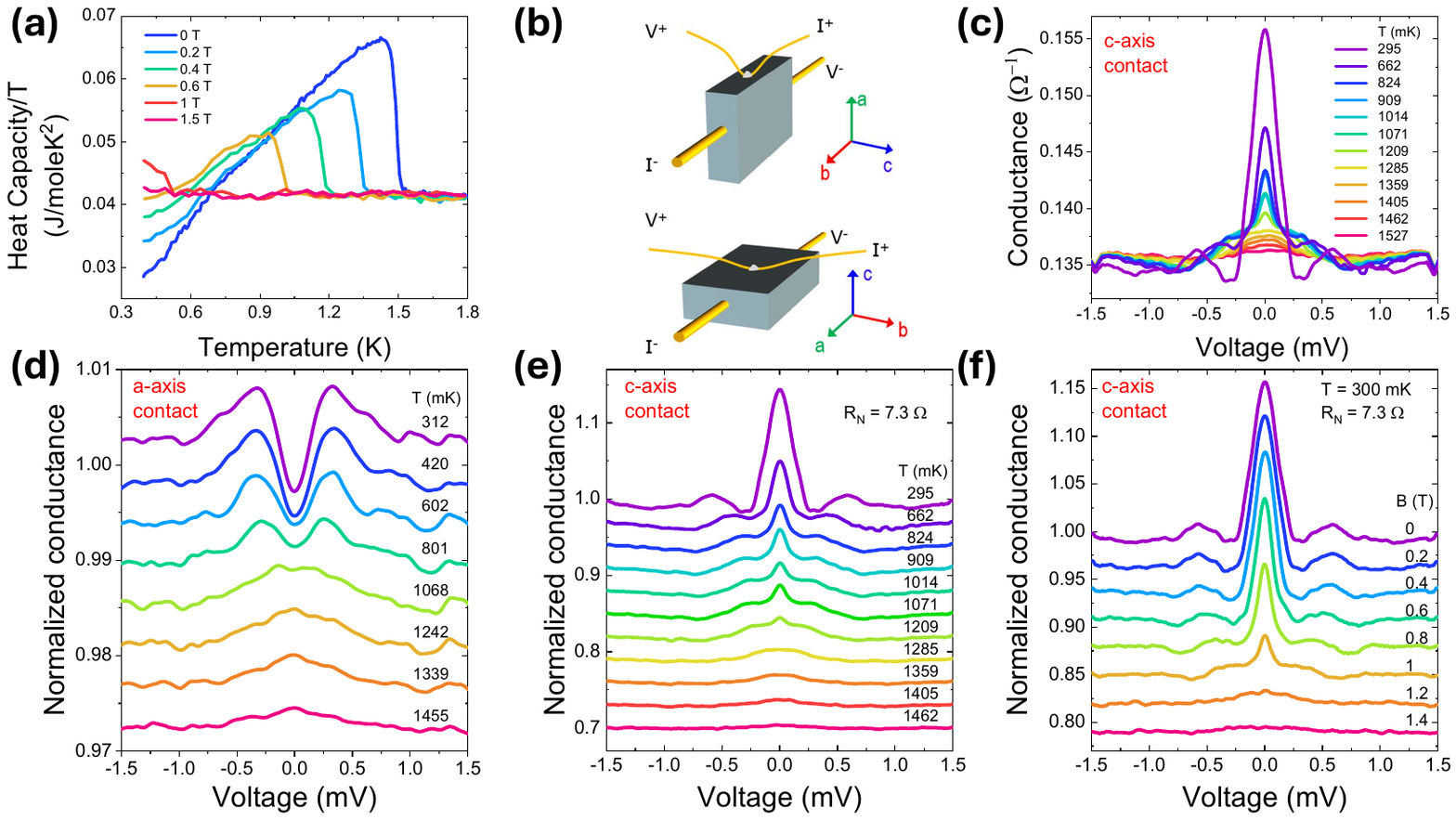}
    \vspace{-2.5cm}
    \caption{
\textbf{Directional point-contact spectroscopy of Sr$_2$RuO$_4$.}
(a) Heat capacity of a SRO single crystal as a function of temperature and magnetic field.
(b) Geometry of the soft point contacts on the [100] and [001] surfaces.
(c) Experimental differential conductance for a c-axis contact as a function of voltage at different temperatures.
(d) Normalized conductance measured using an Ag-paste contact with current injected perpendicular to the basal plane ([001] direction) at different temperatures.
(e) Corresponding conductance for an in-plane injection ([100] direction) at different temperatures.
(f) The same as in (e) but at different in-plane magnetic fields up to $1.4 $T.}
\label{fig:expfig1}
\end{figure*}

In this manuscript, we investigate the Andreev-reflection spectroscopic properties of Sr$_2$RuO$_4$, uncovering a striking reversal of this behavior.  Sr$_2$RuO$_4$ is a prototypical unconventional superconductor that, since its discovery~\cite{Maeno1994}, has remained a central topic in condensed matter physics for nearly three decades. Despite extensive experimental and theoretical efforts, the symmetry of its superconducting order parameter and the microscopic pairing mechanism remain unresolved~\cite{Kivelson2020,maeno2024JPSJ,maeno2024NatPhys}. A broad consensus has emerged that superconductivity in Sr$_2$RuO$_4$ is non-$s$-wave and most likely of spin--singlet--type~\cite{pustogow2019Nature}. 
However, the detailed momentum dependence of the superconducting gap, especially the existence and orientation of nodes, remains a subject of intense debate. Beyond the ongoing controversy surrounding time-reversal symmetry breaking~\cite{Luke1998,Xia2006,Benhabib2021,li2025magfield,Li2021,Li2022,Mazzola2024,Fittipaldi2021,Shiroka2012,Wu20205}, a central unresolved issue is whether the superconducting gap features vertical line nodes, horizontal line nodes, or a mixture of both. Experimental results to date have yielded conflicting interpretations~\cite{Izawa2001,Deguchi2004,Iida2020,Hassinger2017,Wu2020,Landaeta2024}.
From a theoretical standpoint, a symmetry-protected horizontal node at $k_z = 0$ would be somewhat unexpected in a quasi-two-dimensional material such as Sr$_2$RuO$_4$, since it would imply suppressed intralayer pairing. 
Resolving which type of nodes are present is therefore crucial for narrowing down the possible superconducting states in Sr$_2$RuO$_4$. 

Previous tunneling experiments have provided valuable spectroscopic data, but they were generally unable to resolve the complete directional channels of the Andreev contributions~\cite{Laube2000,Kashiwaya2011}. Moreover, these measurements were mostly interpreted using models assuming spin-triplet pairing with odd-parity momentum, which do not account for the possible multiorbital character of the superconducting order parameter~\cite{Yada2014}.

To address these questions, we study how the presence or absence of Andreev bound states (ABS) along different crystallographic directions constrains the symmetry and orbital structure of the superconducting order parameter. We show that inter-orbital pairing naturally generates horizontal nodal lines in the gap and can account for a node at $k_z=0$ in Sr$_2$RuO$_4$, consistent with the orbital composition of its electronic states.
Directional Andreev-reflection (AR) spectroscopy provides a sensitive probe of this scenario because ABS formation depends on the momentum and orbital structure of the pairing~\cite{Kashiwaya_2000}. Unlike the typical behavior of quasi--two--dimensional unconventional superconductors, where in-plane edges show enhanced in--gap spectral weight and out--of--plane conductance is suppressed, Sr$_2$RuO$_4$ displays the opposite trend, which we attribute to the inter-orbital nature of the superconducting state.
Edge- and surface-sensitive AR measurements reveal strong in-gap ABS at the top surface for both even- and odd-parity inter-orbital pairing, while lateral edge modes remain largely gapped. This directional asymmetry provides a fingerprint of inter-orbital superconductivity and is consistent with a gap containing a horizontal line node.
Combining \textit{ab initio} calculations with tight-binding models, including explicit Sr$_2$RuO$_4$/Ag interfaces, we demonstrate the key role of inter-orbital correlations in shaping the ABS and AR conductance. An effective model with a horizontal line node reproduces the main experimental features, linking directional Andreev-reflection spectroscopy to inter-orbital pairing in Sr$_2$RuO$_4$.

\textit{Experimental findings.--}
Very-high-quality single crystals of Sr$_2$RuO$_4$ were grown at St Andrews University using the floating-zone method in an infrared image furnace \cite{Curran2011,Bobowski2019}. We characterized them by various standard techniques, such as Energy Dispersive X-ray spectroscopy (EDX) and mapping, as well as TEM imaging down to the atomic level (see Supplemental Material). All of these analyses showed high perfection of the crystal lattice, and none of them indicated the presence of impurities.
A superconducting phase with critical temperature $T_c = 1.49 \pm 0.02\,\mathrm{K}$ was identified through heat-capacity measurements performed under an applied magnetic field, as shown in Fig. 1~(a).
The superconducting features vanish when the in-plane magnetic field reaches approximately $1.5$ T. Both the critical temperature and the critical field, as well as the temperature dependence of the zero-field heat capacity, are in excellent agreement with previously reported results for the highest-quality samples in the literature \cite{NishiZaki2000,Li2021}. \\
The crystals were then cut along the (100), (110) and (001) planes using a specialized goniometer head that is compatible with both the X-ray Laue back-reflection diffraction system used for crystal orientation and the diamond wire saw used for precise cut. This "Stuttgart method" allows for cutting along any desired crystallographic plane with an accuracy often better than 0.1$^{\circ}$. \cite{stuhlhofer2002}.
The mirror-like crystal surfaces obtained with this method have sizes in the millimeter range and are thus ideal for fabricating ``soft'' point contacts with small drops of Ag paint, a technique we first introduced more than 25 years ago \cite{Daghero_SUST.23}. \\
A schematic drawing of this way of realizing the point contact using a tiny drop (diameter about 50 - 100 $\mu$m) of silver-based conductive paint to which a voltage and a current electrode of the junction are connected is shown in Fig. 1(b). 
We subsequently carried out point-contact Andreev-reflection spectroscopy (PCARS) by measuring the $I-V$ curves of the NS junctions and numerically differentiate them to obtain the $dI/dV$ vs. $V$ curves in different geometrical configurations of the contacts (see Fig. 1(b)). This  enables selective access to distinct crystal terminations within planes perpendicular to the crystallographic axes defined by symmetry, i.e. directional PCARS \cite{Daghero2011,Tortello2010}. 
It is very important to note that, unlike the classic way to obtain NS junctions (needle-anvil technique) where a very sharp metal tip presses on the surface of the sample, in our case the soft contact does not penetrate the sample and, therefore, allows the correct directionality of the electron injection and avoids spurious pressure effects.
\\
For differential conductance measured along the c~axis, we observed a pronounced zero-bias conductance peak that vanishes at the superconducting transition temperature [Fig. 1(c)]. In addition to this dominant zero-bias feature, two weaker bump structures appear at approximately 0.6 mV and shift toward zero energy as the temperature increases toward the transition. The progressive merging of these satellite features with the zero-bias peak results in a substantial broadening of the low-bias conductance close to the transition temperature.
For in-plane PCARS, the normalized conductance shows a suppression of in-gap spectral weight, accompanied at low temperatures by two finite-bias peaks [Fig. 1(d)]. These features appear at voltage amplitudes (0.32-0.33 mV) lower than those observed for the broad bumps of the c-axis conductance and shift toward zero bias as the temperature approaches the superconducting transition. We observed the same spectroscopic features in several contacts along the (001), (100) and (110) directions. Some examples showing the reproducibility of these results can be found in Supplemental Material. The superconducting origin of the structures in the normalized conductance along the c-axis is evidenced by their temperature behavior shown in Fig. 1(e). This is further confirmed by the response of the 
c-axis normalized conductance to an out-of-plane magnetic field, which fully suppresses all features above the upper critical field of $\sim$ 1.4 T as highlighted in Fig. 1(f). At the increase of magnetic field the zero-bias peak progressively reduces in intensity and the finite-bias bumps progressively shift toward lower energy, finally merging in a single bump at $H \sim$ 1.2 T. However, the behavior of these structures, as well as that of the dips separating the zero-bias peak from the finite-bias bumps, is qualitatively very different if superconductivity is suppressed by temperature or magnetic field. 
Further details on the experimental methods and instrumentation for heat capacity and PCARS measurements can be found in Supplemental Material.
\\
Similar spectroscopic features were observed in the few Andreev-reflection experiments on  Sr$_2$RuO$_4$ previously carried out \cite{Laube2000,Fogelstro_PhysicaB_2000,Raychaudhuri,Kashiwaya2011,WangPRB2015} but, except in Ref.\cite{Raychaudhuri}, with considerable uncertainty about the directionality of the contact.

\begin{figure*}[!t]
    \centering
    \includegraphics[width=1.00\linewidth]{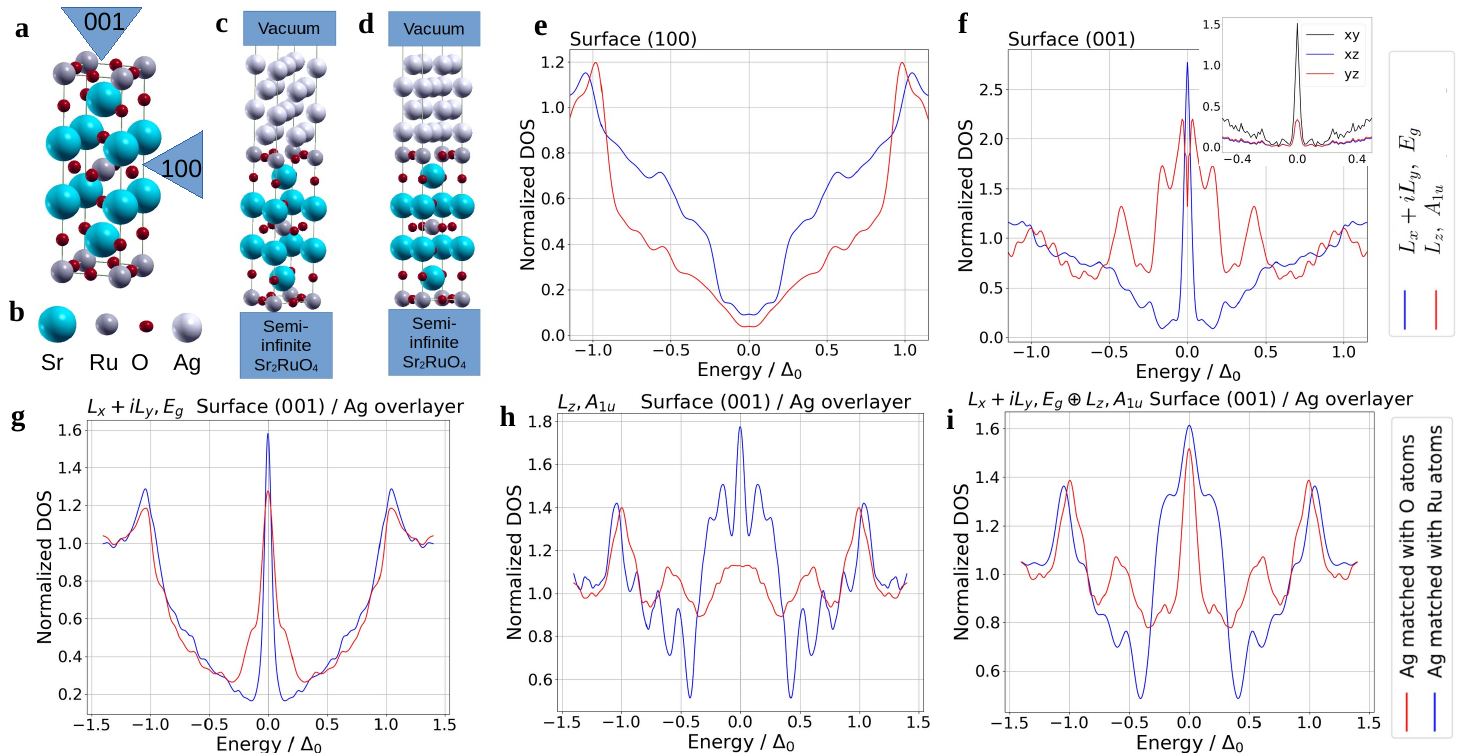}
    \caption{
\textbf{Quasiparticle Density of States on Different Surface Structures.} 
(a) Crystal structure of Sr$_2$RuO$_4$ highlighting the (100) and (001) surfaces. 
(b) Atomic species used in panels (a) and (c-d). 
(c-d) Surface terminations with Ag overlayers on semi-infinite Sr$_2$RuO$_4$: 
(c) Ag atoms aligned with surface O atoms; 
(d) Ag atoms aligned with surface Ru atoms. 
(e) Quasiparticle Density of States (QPDOS) for the (100) surface assuming inter-orbital $E_g$ 
and $A_{1u}$ pairing symmetries. No Andreev bound states are present. 
(f) QPDOS for the (001) surface with the same pairing symmetries. For $E_g$ pairing, a pronounced, narrow zero-energy peak is observed; the inset shows the orbital-resolved contributions to this peak. For $A_{1u}$ pairing, dispersive Andreev bound states are present, accompanied by a weaker zero-energy peak. 
(g-i) QPDOS for the topmost Ag layer, assuming 12 Ag overlayers as in structures (c) and (d): 
(g) $E_g$ pairing results in spectral broadening due to the Ag overlayer; 
(h) $A_{1u}$ pairing produces a peak-dip structure; 
(i) Mixed $E_g$-$A_{1u}$ pairing yields similar peak-dip structures consistent with experiment.
These results, obtained from density functional Dirac-Bogoliubov-de Gennes calculations, indicate that the observed anisotropic Andreev bound states require inter-orbital pairing, realized through $E_g$, $A_{1u}$, or mixed symmetry components.}
\label{fig:dft_result}
\end{figure*}
\begin{figure*}[t!]
    \centering
    \includegraphics[width=17.5cm]{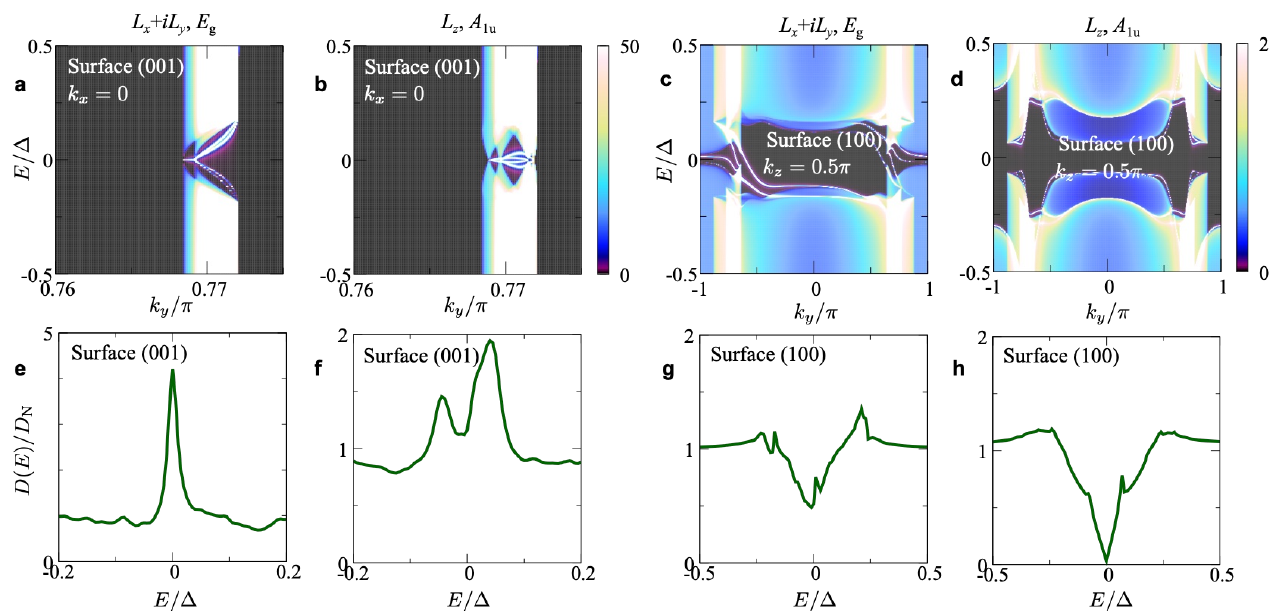}
    \caption{\textbf{Surface density of states.} (a,c,e,g) $L_x+iL_y$ $E_g$ and (b,d,f,h) $L_z$ $A_\mathrm{1u}$ pairings.
    (a,b) Surface density of states at the (001) surface for $E$ and $k_y$ space at $k_x=0$.
    (c,d) Surface density of states at the (100) surface for $E$ and $k_y$ space at $k_x=0.5\pi$.
    (e,f) Surface density of states at the (001) surface as a function of $E$.
    (g,h) Surface density of states at the (100) surface as a function of $E$.
    We choose the parameters as $\Delta=1.8\times10^{-4}t$ and $\delta=10^{-4}\Delta$ for the spectral function broadening in panels (a,b,c,d,g,h), $\delta=10^{-2}\Delta$ in panels (e,f), and $\delta=10^{-3}\Delta$ in panels (g,h).}
    \label{fig:SABSs}
\end{figure*}%

\textit{Surface Andreev bound states by density functional theory and tight-binding approaches.--}
In Sr$_2$RuO$_4$, the low-energy electronic structure is dominated by the $t_{2g}$ orbitals $d_{xz}$, $d_{yz}$, and $d_{xy}$, which allow not only intra-orbital but also inter-orbital superconducting pairing \cite{maeno2024JPSJ,Ramires2019,SuhPRR2020,Fukaya2018,Fukaya2019}. Inter-orbital pairing occurs between electrons in different orbitals, such as $d_{xz}-d_{yz}$ or $d_{xz/yz}-d_{xy}$, and can appear in either spin-singlet or spin-triplet channels. The parity of the pairing can be even or odd, depending on the momentum and orbital structure, and spin--orbit coupling can mix these channels to produce multi-component order parameters. 

In order to gain microscopic insight into how different surface orientations and inter-orbital pairing symmetries affect the low-energy quasiparticle spectrum, we start by performing density-functional Dirac-Bogoliubov-de Gennes calculations using the screened Korringa-Kohn-Rostoker Green's function method \cite{Capelle1999,Csire2018} (see Supplemental Material).
This approach enables us to incorporate the realistic Fermi surface, orbital composition, and spin-orbit coupling of Sr$_2$RuO$_4$, while simultaneously accounting for semi-infinite geometries to model surfaces and overlayers.
In this way, superconducting pairing is treated on the same footing as the underlying ab-initio electronic structure.
Figure~\ref{fig:dft_result} summarizes the results for the bare (100) lateral and (001) basal-plane surfaces (Fig.~\ref{fig:dft_result}(a)), while also showing the effects of coherent Ag overlayers placed in different registry variations relative to the Sr$_2$RuO$_4$ basal-plane surface (Figs.~\ref{fig:dft_result}(c--d)).

By comparing the quasiparticle density of states across different symmetry channels and surface terminations, we can identify how the presence or absence of zero-energy Andreev bound states depends on the orbital structure of the pairing symmetry.
In particular, the inter-orbital $E_g$ ($d_{xz/yz}-d_{xy}$) and $A_{1u}$ ($d_{xz}-d_{yz}$) are representative pairings as they produce distinctive low-energy features for the basal-plane surface, whereas no significant in-gap states are observed for the lateral surface as can be seen in Fig.~\ref{fig:dft_result}(e--f).
For the basal-plane surface, the inter-orbital $A_{1u}$ pairing gives rise to a richer and more broadened in-gap structure, not limited to a single zero-energy peak, in contrast to the inter-orbital $E_g$ pairing, which exhibits a very sharp zero-energy peak.
These results, therefore, highlight the role of inter-orbital pairing to account for the observed surface-dependent spectra.
Another distinctive feature of the emergent Andreev bound states lies in their pronounced orbital character.
As illustrated in the inset of Fig.~\ref{fig:dft_result}(f), for inter-orbital $E_g$ pairing the orbital-resolved spectral weight of the zero-energy peak is predominantly of $d_{xy}$ character. Since this electronic state is mainly confined to the plane, it is expected to couple weakly to a conducting channel oriented largely out of plane. This provides a possible explanation for the absence of the peak in scanning tunneling spectroscopy measurements. In contrast, directional Andreev-reflection spectroscopy via point-contact techniques can promote orbital mixing at the interface with the metallic nanoscopic island, thereby generating a finite spectral weight and enabling the detection of the zero-energy feature.
As illustrated in the inset of Fig.~\ref{fig:dft_result}(f), for inter-orbital $E_g$ pairing the orbital-resolved spectral weight of the zero-energy peak is predominantly of $d_{xy}$ character. Since this electronic state is mainly confined to the plane, it is expected to couple weakly to a conducting channel oriented largely out of plane. This provides a possible explanation for the absence of in-gap peak in the conductance probed by scanning tunneling spectroscopy~\cite{Suderow_2009}. In contrast, directional Andreev-reflection  spectroscopy via point-contact techniques can promote orbital mixing at the interface with the metallic nanoscopic island, thereby generating a finite spectral weight and enabling the detection of the zero-energy feature.

Indeed, in the experiment the point contact is realized by a small drop of silver paste, which one may approximate by adding several Ag overlayers to investigate their impact on the quasiparticle spectra.
As illustrated in Fig. \ref{fig:dft_result}(c--d), we consider two representative surface reconstructions: one where the first Ag layer is aligned with the surface oxygen atoms, allowing for stronger hybridization, and another where the first Ag layer is aligned with the Ru atoms, resulting in comparatively weaker coupling between the Ag and the substrate.
We include both configurations to reflect the fact that, in the experimental setup, the Ag atoms may not adopt a single well-defined alignment but instead form locally varying arrangements at the interface, which can influence the degree of hybridization and spectral broadening.
This behavior is directly observed in Fig.~\ref{fig:dft_result}(g), where 12 Ag overlayers are placed on the (100) surface assuming inter-orbital $E_g$ pairing.
The zero-energy peak is always broadened in comparison to the bare surface, and the broadening is stronger when the Ag atoms are aligned with the oxygen atoms because of stronger hybridization.
A similar qualitative behavior is found for the inter-orbital $A_{1u}$ pairing in Fig.~\ref{fig:dft_result}(h), where the in-gap Andreev bound states become increasingly broadened if the Ag atoms and O atoms are aligned.
Since inversion symmetry is broken at the surface, particularly in the presence of Ag overlayers, mixing of even- and odd-parity pairings becomes allowed by symmetry at the surface.
This scenario is shown in Fig.~\ref{fig:dft_result}(i), where the pairing contains equal contributions of $E_g$ and $A_{1u}$ components, exhibiting a combined broadening behavior similar to that of the individual pairings.

Our ab-initio analysis provides a detailed picture of the electronic structure and hints at the mechanisms underlying the low-energy Andreev spectra within a realistic surface and interface setup. To translate these insights into a more physically intuitive framework, we employ a microscopic three-band model in which the oxygen degrees of freedom are projected out \cite{SuhPRR2020}. This multiband approach allows us to trace how inter-orbital pairing in the band basis generates the anisotropic behavior of the pairing amplitude and to resolve the momentum-dependent dispersion of the Andreev bound states, including the sign change along the $c$-axis. By bridging ab-initio results with this effective model, we gain a clear understanding of the origin of in--gap surface states and the role of inter-orbital pair correlations.


We start by showing that interlayer hybridization naturally generates an anisotropic sign reversal of the pairing amplitude when interorbital pairing is considered \cite{FukayaPRR02022,AndoPRB2022,Autieri25,Ramires2025}, whether in the intra-layer channels ($d_{xz/yz}-d_{xy}$) or ($d_{xz}-d_{yz}$). This mechanism inherently produces surface in-gap Andreev bound states, illustrating how the combination of multiband structure and interorbital pairing shapes the low-energy spectral landscape (see Supplemental Material).
To proceed further, it is useful to introduce the Gellmann matrices ($\hat{L}_i$, generators of the SU(3) algebra, that is a convenient basis for representing the structure of three-orbitals quantum systems (see Supplemental Materials) as well as the Pauli matrices $\hat{\sigma}_{j}$ in spin space. 
The normal state Hamiltonian in the bulk can generally be expressed
\begin{align}
    \hat{H}(\bm{k})&=\sum^{8}_{i=0}\sum^{3}_{j=0}h_{i,j}(\bm{k})\hat{L}_{i}\otimes\hat{\sigma}_{j},
\end{align}%
where $h_{i,j}(\bm{k})$ are the momentum dependent orbital and spin-orbital terms related to electron hybridization and the atomic spin-orbit coupling. Details of the form and the parameterization of $h_{i,j}(\bm{k})$ are reported in the Supplemental Material. 
We now turn our attention to the interlayer hopping processes and examine how the interorbital pairing is modified upon transformation to the rotated band basis. In particular, we focus on the case of $E_g$ pairing. 
Focusing on the spin--orbit--induced interlayer hopping processes between the $xy$ and $xz/yz$ orbitals, described by the operators $\hat{L}_{5}$ and $\hat{L}_{6}$, the effective interlayer Hamiltonian can be expressed as
\begin{align}
    \hat{H}_{\mathrm{eff}}(\bm{k})
    = h_{5,3}(\bm{k}) \hat{L}_{5} \otimes \hat{\sigma}_{3}
    + h_{6,3}(\bm{k}) \hat{L}_{6} \otimes \hat{\sigma}_{3}.
\end{align}
The momentum-dependent coefficients are given by
\begin{align}
    h_{6,3}(\bm{k})
    &= 8 \gamma^{\mathrm{so}}_{56,z}
    \sin\frac{k_z}{2}
    \sin\frac{k_x}{2}
    \cos\frac{k_y}{2}, \\
    h_{5,3}(\bm{k})
    &= -8 \gamma^{\mathrm{so}}_{56,z}
    \sin\frac{k_z}{2}
    \cos\frac{k_x}{2}
    \sin\frac{k_y}{2}.
\end{align}

Performing an appropriate unitary transformation to the band basis (see Supplemental Material), one finds that the resulting intraband $E_g$ pairing acquires a nontrivial momentum dependence of the form
\begin{align}
    \Delta(\bm{k})
    = \Delta \,
    \frac{h_{6,3}(\bm{k}) + i h_{5,3}(\bm{k})}
    {\sqrt{h_{6,3}^2(\bm{k}) + h_{5,3}^2(\bm{k})}}\, ,
\end{align}
with $\Delta$ indicating the amplitude of interorbital pairing.
This expression shows that the pairing develops a nodal component along the $k_z$ direction. As a consequence, the superconducting gap may undergo a sign change as a function of $k_z$, leading to the formation of Andreev in-gap bound states for a surface perpendicular to the $c$-axis. 
This behavior is also obtained when considering the other types of inter-orbital pairings, either with $A_{1u}$ or B$_{1g}$ symmetry.


We then consider the momentum-resolved and integrated density of states along the $z$- and $x$-direction for the $E_g$ and $A_{1u}$ states (Figs.~\ref{fig:SABSs} a-h) for the model analysis. 
We compute the surface Green's function for semi-infinite superconductors and evaluate the surface density of states ~\cite{Umerski,Takagi20}(see Supplemental Material).

For the inter-orbital $E_g$ state, zero-energy flat bands appear for the [001] surface, as shown in Fig.~\ref{fig:SABSs}(a), leading to pronounced zero energy peak in the density of states (Fig.~\ref{fig:SABSs}(e)).
In contrast, for the interorbital $A_{1u}$ state, the surface Andreev bound states exhibit a weak splitting [Fig.~\ref{fig:SABSs}(b)], and consequently two peaks can be resolved close to the zero energy in the integrated density of states [Fig.~\ref{fig:SABSs}(f)].
Along the $x$ direction, we find dispersive edge modes for the inter-orbital $E_g$ pairing [Fig.~\ref{fig:SABSs}(c)], whereas a gapped structure without surface Andreev bound states is obtained for the interorbital $A_{1u}$ pairing [Fig.~\ref{fig:SABSs}(d)].

The momentum-space analysis explains why the $E_g$ pairing exhibits a sharp zero-bias peak originating from the presence of flat bands, while the $A_{1u}$ pairing supports broader features close to zero energy due to the small splitting of the flat bands.
This is a general behavior that is expected for similar types of inter-orbital pairing that involve the $xz$-$yz$ orbitals.
For the in-plane [100] termination, the $E_g$ superconducting state hosts in-gap Andreev bound states that are strongly momentum dependent and occurring in small portion of the momentum space [Fig.~\ref{fig:SABSs}(c)]. This behavior leads to a partial suppression of the in-gap spectral intensity [Fig.~\ref{fig:SABSs}(g)].
By contrast, the $A_{1u}$ pairing exhibits a fully gapped spectral function [Fig.~\ref{fig:SABSs}(d)] that lead to a pronounced suppression of the in-gap spectral intensity of the integrated density of states [Fig.~\ref{fig:SABSs}(h)]. This analysis confirms the anisotropic behaviour of the Andreev bound states and highlight the character of the Andreev states in momentum space.

\textit{Model for point-contact Andreev conductance --}
The above analysis shows that inter-orbital pairing, no matter whether in the $E_g$ and $A_{1u}$ channels, generates a superconducting pairing amplitude that changes sign along the $c$ axis, giving rise in turn to robust zero-energy peaks or broader subgap structures in the QPDOS along on the $(001)$ surface, and no subgap states on the  $(100)$ surface.  This is perfectly compatible with the results of point-contact spectroscopy measurements, where zero-bias peaks were observed in c-axis contacts (i.e. when probing the $(001)$ surface) and not in a-axis contacts (i.e. when probing the $(100)$ surface).  
Andreev-reflection spectra are actually the differential conductance of the normal metal/superconductor junction plotted as a function of the bias voltage, i.e. $dI/dV$ vs. $V$, and cannot be directly compared to the QPDOS. However, we can construct an effective model that grasps the key feature of both these interorbital pairings, i.e. a sign change in the superconducting pairing along the $c$ axis, and see if it can explain the experimental findings -- at least in a qualitative way. We will thus assume that the superconducting pairing has a simple dependence $\Delta = \Delta_0 \sin (k_z)$. To calculate the theoretical conductance curves, we will use a generalization of the well-known Blonder-Tinkham-Klapwijk model (BTK) \cite{BTK} to the case of an anisotropic superconductors, and particularly suited to describe materials with a cylindrical Fermi surface, as is the case of strontium ruthenate \cite{Yamashiro_PhysRevB.56}. An effective, cylindrical Fermi surface (that mimics the $\beta$ or $\gamma$ sheets of the actual one \cite{Firmo2013}) with the pairing amplitude on top is shown in Figure 4: the color of the surfaces indicates the different sign of the order parameter above and below the basal plane. The $\sin(k_z)$ dependence gives rise to a node line at $k_z=0$ and a sign change above/below the basal plane, which is expected to result in Andreev bound states \cite{TK95}  {only} when the current is injected along the $c$ axis, thus giving rise to zero-energy conductance peaks.

\begin{figure}[t!]
\vspace{-1cm}
    \centering
     \includegraphics[width=8.5cm]{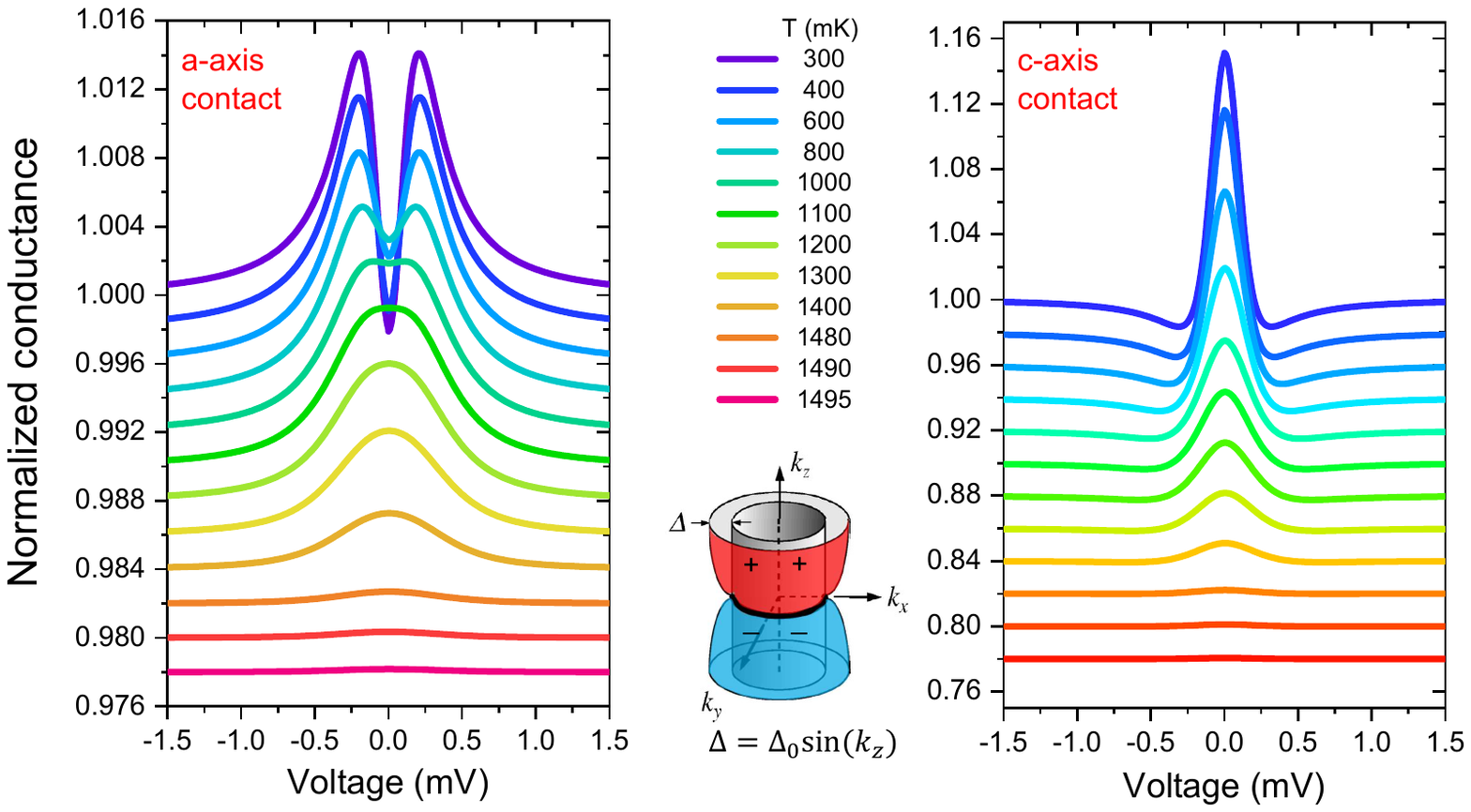}
     \vspace{-1cm}
     \caption{\textbf{Andreev spectra calculated with the effective generalized BTK model}. Centre: sketch of the cylindrical Fermi surface with the order parameter $\Delta = \Delta_0(T) \sin (k_z)$ on top. The color refers to the sign of the pairing. Left and right panels report the spectra calculated for a-axis and c-axis contacts, respectively, at different temperatures. }
    \label{fig:pcars}
\end{figure}%
Figure \ref{fig:pcars} shows the theoretical spectra calculated for a-axis contacts (left) and c-axis contacts (right) as a function of temperature (see Supplemental material for details). In addition to the gap amplitude $\Delta_0$, the model contains a dimensionless parameter $Z$ (proportional to the height of the potential barrier at the normal/superconductor interface) and can be made more adherent to a real experiment by introducing a broadening parameter $\Gamma$ which is actually the imaginary part of the energy \cite{Plecenik1994,srikanth92,Daghero_SUST.23}. The spectra in Figure 4 were all calculated with the same parameters, i.e. $Z= 0.8$, $\Delta_0= 0.4\,\mathrm{meV}$ and $\Gamma = 0.15\,\mathrm{meV}$. To account for the different temperature, the theoretical curves at $T=0$ were convoluted with the Fermi function. Moreover, a BCS-like temperature dependence was assumed for the gap amplitude, i.e. $\Delta_0(T) = \Delta_0 (T=0)\tanh(1.74 \sqrt{T_c/T-1})$.

The model successfully captures the key characteristics of the experimental spectra. For ab-plane contacts, it reproduces the two symmetric conductance peaks around 0.4 meV, separated by a minimum at zero bias. In the case of c-axis contacts, the model accurately reflects the presence of a pronounced zero-bias peak surrounded by two shallow minima. Moreover, with the chosen parameters, the signal amplitudes along the different crystallographic directions closely match the experimental observations.

\textit{Conclusions --} In this work we have investigated the spectroscopic signatures of superconductivity in Sr$_2$RuO$_4$ by analyzing the directional dependence of Andreev bound states at edges and surfaces. Our results reveal an unconventional anisotropy of the AR spectra, characterized by pronounced in-gap states at  surfaces perpendicular to the out-of-plane direction and suppressed spectral weight at lateral surfaces. This behavior contrasts with the conventional expectation for quasi-two-dimensional unconventional superconductors, where planar edges typically host the most prominent Andreev bound states.

We have shown that this reversed anisotropy can arise naturally from the inter--orbital character of the superconducting pairing. Both even- and odd--parity inter--orbital pairing channels produce robust (001) surface Andreev bound states while leaving lateral edge modes largely suppressed, providing a direct probe of the orbital structure of the pairing state. Inter--orbital pairing also provides a natural mechanism for the emergence of a horizontal line node in the superconducting gap, which plays a central role in reproducing the directional features of the tunneling conductance in our effective model. Moreover, our analysis indicates that surface reconstructions can induce mixed--parity pairing at the termination of the material. In particular, while the odd-parity $A_{1u}$ configuration is unlikely to be realized in the bulk, it can appear near the surface where the reduced crystalline symmetry allows for mixed-parity contributions. This effect contributes to a broader distribution of Andreev spectral weight with more dispersive features, highlighting the role of ($d_{xz}$--$d_{yz}$) inter-orbital hybridization in shaping the surface spectra. 

We also note that these directional Andreev features are consistent with a multicomponent superconducting order parameter originating from the multiorbital nature of the pairing in Sr$_2$RuO$_4$ as demonstrated for the case of the $E_g$ pairing. In contrast, the presence of vertical nodes would be difficult to reconcile with the observed anisotropy in the conductance.

By combining \textit{ab initio} calculations with tight-binding modeling and explicit treatment of Sr$_2$RuO$_4$/Ag interface reconstructions, we have shown that the main AR conductance features can be captured by an effective model incorporating inter-orbital correlations, the horizontal line node, and surface-induced mixed parity. However, the present spectroscopic evidence cannot provide direct information on the presence or absence of time-reversal symmetry breaking, as for instance both chiral and non-chiral inter-orbital pairing channels of the $E_g$ type would produce qualitatively similar directional Andreev spectroscopic features. 

Overall, our results highlight the crucial role of orbital degrees of freedom, inter-orbital hybridization, and interface effects in shaping the low-energy spectroscopic response of Sr$_2$RuO$_4$. They establish a consistent framework linking inter-orbital pairing, multicomponent superconducting order, the directional asymmetry of Andreev bound states, the possible presence of a horizontal line node, and surface-induced mixed-parity effects. While further experimental and theoretical studies are needed to resolve the question of time-reversal symmetry breaking, directional Andreev-reflection spectroscopy provides a powerful probe of the orbital and nodal structure of the superconducting order parameter.


\noindent $^{*} $G.C.,Y.F. and R.S.G. contributed equally to this work.
M.C. and G.C. acknowledge partial support from
NRRP MUR project PE0000023-NQSTI. M.C. acknowledges support by Italian Ministry of University and Research (MUR) PRIN 2022 under the Grant No. 2022LP5K7 (BEAT). R.S.G. and D.D. acknowledge the support of the Italian Ministry of University and Research (MUR) PRIN 2017 under Grant No. 2017Z8T55B (Quantum2D). R.S.G. expresses his deepest gratitude to Oleg Dolgov, who sadly passed away, for his long-standing friendship and for being the first to suggest the simple order-parameter symmetry used here in our effective model.
Y.\,F.\, acknowledges financial support from the Sumitomo Foundation and JSPS with Grants-in-Aid for Scientific Research (KAKENHI Grants No.\ 26K17096), and the calculation support from Okayama University.  
Y.\, T.\, acknowledges financial support from JSPS with Grants-in-Aid for Scientific
Research (KAKENHI Grants Nos. 23K17668, 24K00583, 24K00556,
24K00578, 25H00609 and 25H00613).

\bibliography{biblio}
\onecolumngrid


\end{document}


\title{Supplemental Material: Directional Andreev-Reflection Signatures of Inter-Orbital Pairing in Sr$_2$RuO$_4$}

\author{G. Csire*}
\affiliation{CNR-SPIN, c/o Universit\`a di Salerno, IT-84084 Fisciano (SA), Italy}

\author{Y. Fukaya*}
\affiliation{Faculty of Environmental Life, Natural Science and Technology, Okayama University, 700-8530 Okayama, Japan}

\author{M. Cuoco}
\affiliation{CNR-SPIN, c/o Universit\`a di Salerno, IT-84084 Fisciano (SA), Italy}

\author{Y. Tanaka}
\affiliation{Department of Applied Physics, Nagoya University, Nagoya 464-8603, Japan}

\author{R.K. Kremer}
\affiliation{Max-Planck Institute for Solid State Research, Stuttgart, Germany}


\author{A.S. Gibbs}
\affiliation{University of St Andrews, St Andrews, Scotland, UK}

\author{G.A. Ummarino}
\affiliation{Department of Applied Science and Technology, Politecnico di Torino, Torino, Italy}

\author{D. Daghero}
\affiliation{Department of Applied Science and Technology, Politecnico di Torino, Torino, Italy}

\author{R.S. Gonnelli*}
\affiliation{Department of Applied Science and Technology, Politecnico di Torino, Torino, Italy}

\begin{abstract}
This Supplemental Material provides methodological details supporting the main text. We start with a section which describes in more detail the experimental methods employed in the work, presents some additional characterization of the samples and shows the reproducibility of Andreev-reflection results. We then describe the first-principles framework used to study superconductivity in Sr$_2$RuO$_4$, based on the Kohn--Sham--Dirac--Bogoliubov--de Gennes formalism combined with the Korringa--Kohn--Rostoker Green's-function method, which captures the realistic electronic structure and spin--orbit coupling while treating superconducting pairing. We also outline the calculation of the surface and planar-edge density of states in the three-dimensional model, analyze the pair potential in the band basis, and summarize the Blonder--Tinkham--Klapwijk approach used to compute point-contact spectra at normal-metal--superconductor interfaces.
\end{abstract}
\maketitle

\section{S1: Experimental methods, further characterization and reproducibility}

SEM EDX analysis was performed by scanning both the lateral side (ac or bc plane) and the top surface (ab plane) of the single crystals and collecting the Ru L$_\alpha$, Sr K$_\alpha$, and O K$_\alpha$ maps. The crystals are very homogeneous and there are no indications of the presence of a Ru-rich phase with T$_c$=3 K. The point spectra and subsequent chemical analysis show no trace of spurious elements, contaminants, or inclusions, and the stoichiometry is close to ideal. 

TEM diffraction patterns confirm that the Sr$_2$RuO$_4$ crystallites are generally very pure. In very few cases an intergrowth with Sr$_3$Ru$_2$O$_7$ lamellas was observed but there was no evidence of crystallites of this phase. Figure S1 shows a TEM diffraction image at atomic resolution of a crystallite with regular Sr$_2$RuO$_4$ structure evidenced by the white dashed square and a Sr$_3$Ru$_2$O$_7$ lamella by the yellow dashed one. 

Heat capacity curves as function of temperature and magnetic field were recorded using a Quantum Design PPMS instrumentation equipped with a He3 insert and a superconducting magnet. 

Point-contact Andreev-reflection spectroscopy (PCARS) measurements were performed by realizing small point-like Ag-paste contacts on the freshly-cleaved surface of  Sr$_2$RuO$_4$ crystals cut previously along different crystallographic directions, as described in the main text. The differential conductance spectra $dI/dV$ vs. $V$ --which result from the parallel of several nanoscopic contacts-- were obtained by numerical differentiation of the current-voltage curves measured in the pseudo-four-probe configuration. The current was supplied by a high precision ideal current source while the voltage was measured by a nanovoltmeter. The NS junctions thus created were mounted inside the small measuring chamber of a home-made He3 cryostat.

Figure S2 shows the reproducibility of the PCARS results presented in the main text. In particular, in panel (a) the differential conductance of an ab-plane, i.e. a (110) direction, contact is shown. The corresponding normalized conductance is presented in panel (b). Panels (c) and (d) highlight the reproducibility of the PCARS measurements along the c-axis. In both cases, the normalized conductance exhibits a sharp peak at zero bias with characteristics very similar to those shown in Fig. 1 (c), (e), and (f) of the main text. It is important to note that the bias voltages at which the main structures of the curves are visible and the heights of the curve maxima are absolutely similar to those shown in Fig. 1 of the main text. 

\begin{figure} [h]
    \centering
\includegraphics[width=\linewidth]{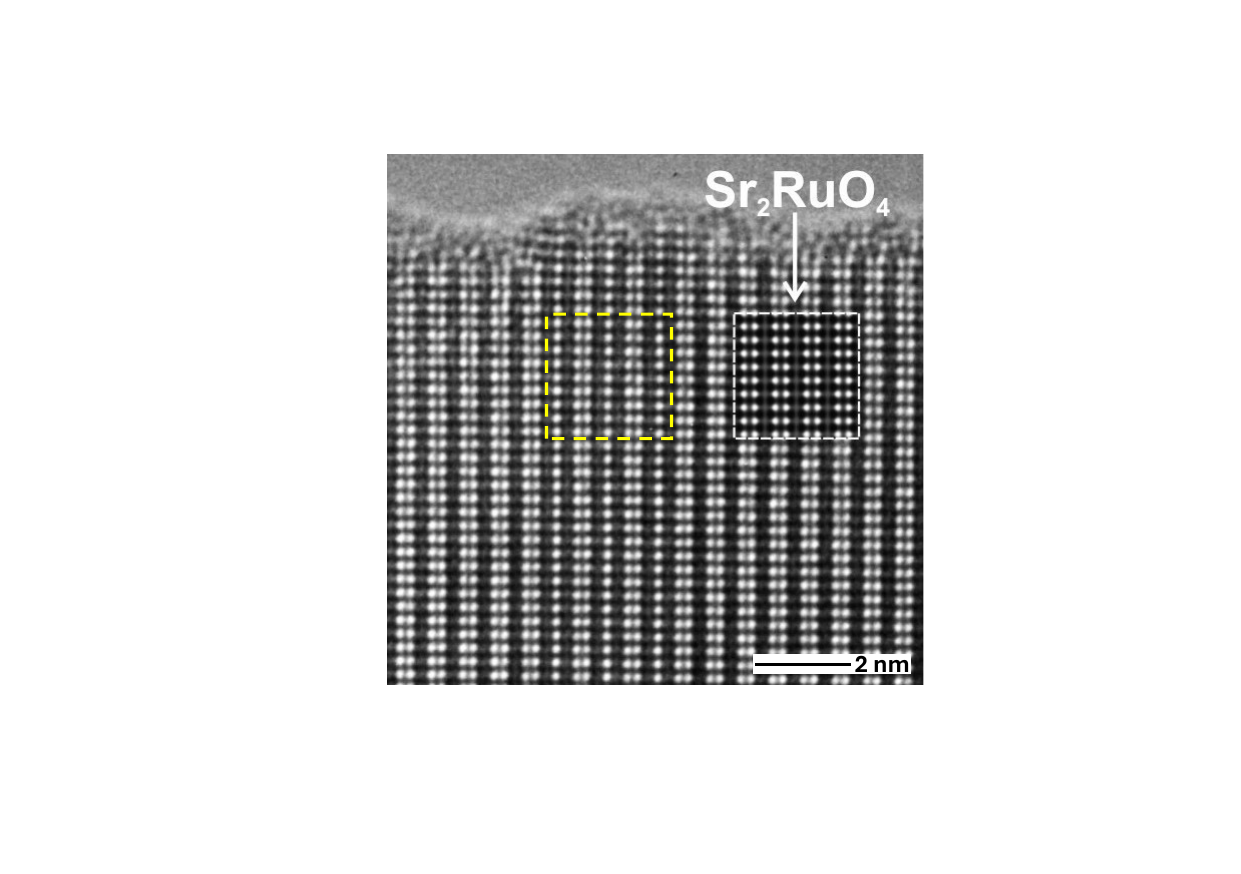}
    \caption{Atomic-resolution TEM image of a Sr$_2$RuO$_4$ single crystal}
    \label{fig:SuppFigTEM}
\end{figure}

\begin{figure*} [t]
    \centering
\includegraphics[width=\textwidth]{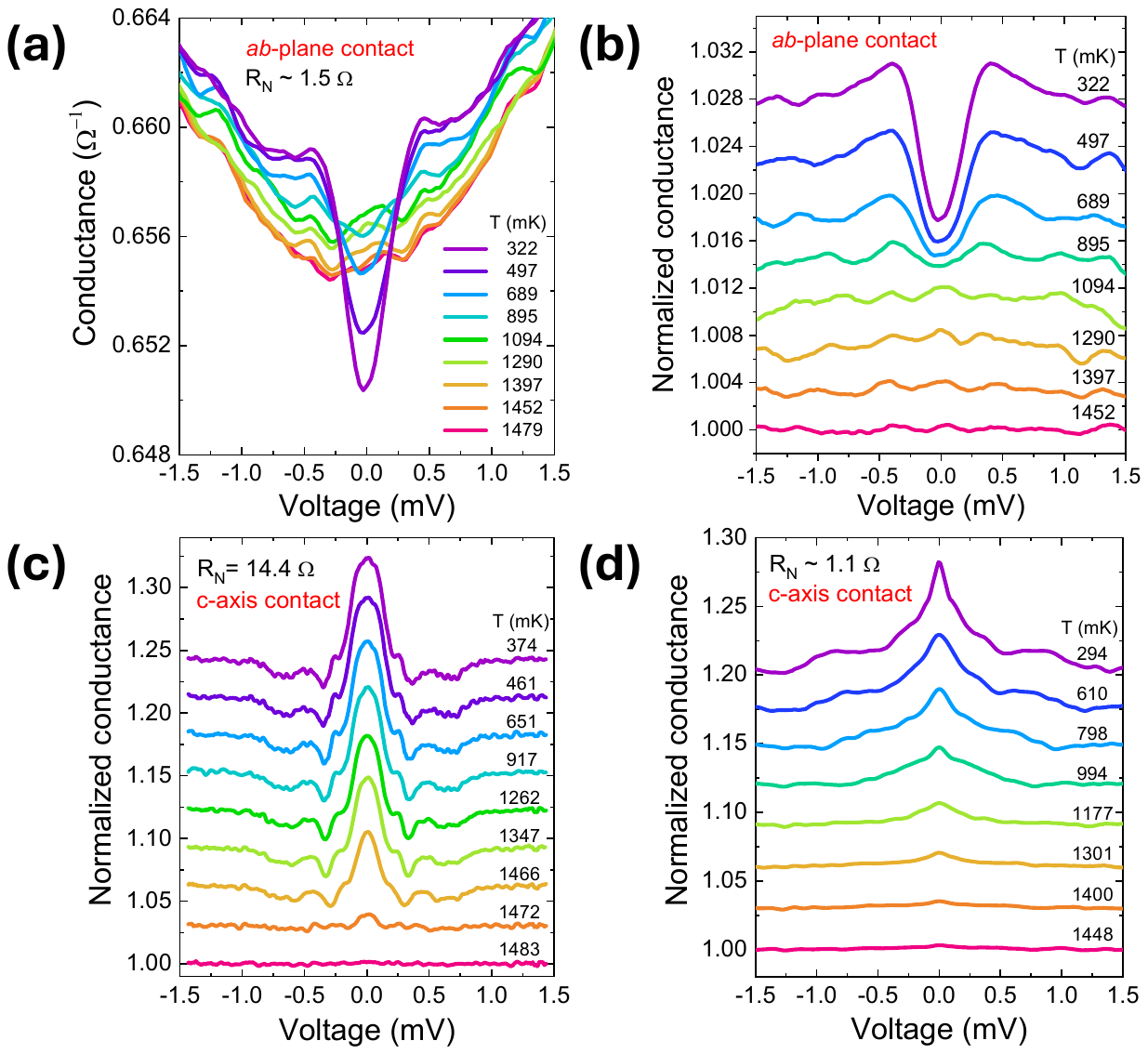}
    \caption{Other examples of directional point-contact Andreev-reflection spectra collected in an ab-plane contact (a) and (b) and in c-axis ones (d) and (e)}
    \label{fig:SuppFigReproduc}
\end{figure*}

\section{S2: First-principles based treatment of superconductivity}

\subsection{Kohn-Sham-Dirac-Bogoliubov-de Gennes equations}

We employ an \emph{ab initio}-based microscopic theory to capture the normal-state electronic structure $-$ including the realistic Fermi-surface geometry, orbital composition, and relativistic effects such as spin-orbit coupling $-$ while treating superconducting pairing on the same footing.
This is based on the relativistic Dirac--Bogoliubov--de~Gennes (DBdG) Hamiltonian which was established by the work of Capelle \emph{et al}.~\cite{Capelle1999} that was later combined with density functional theory in Ref.~\onlinecite{Csire2018}, written in Rydberg units ($\hbar=1$, $m=1/2$, $e^2=2$)
\begin{equation}
H_{\text{DBdG}}=
 \begin{pmatrix}
   H_D & \mathcal D   \\
   \mathcal D^\dagger  & -H_D^\ast
 \end{pmatrix} \, ,
\end{equation}
where
\begin{equation}
 H_D=c {\boldsymbol{\alpha}}{p} + \left( \boldsymbol{\beta}-\mathbb{I}_4 \right) c^2/2+ \left( V_{\text{eff}}(\mathbf r)-E_F \right) \mathbb{I}_4,
\end{equation}
\begin{equation}
  {\boldsymbol \alpha} =
 \begin{pmatrix}
  0 &  {\boldsymbol \sigma} \\
   {\boldsymbol \sigma} & 0
  \end{pmatrix} \, , \qquad
 \boldsymbol \beta=
 \begin{pmatrix}
  \mathbb I_2 & 0 \\
  0 & -\mathbb I_2
  \end{pmatrix} \, ,
\end{equation}
and the effective electrostatic potential $V_{\text{eff}}( \mathbf r)$ obtained using the usual (local density approximation) exchange-correlation energy for normal state electrons, while in general $\mathcal{D}$ contains the pairing potential as kernel
$\mathcal{D}=\int d  \mathbf r' ... \Delta( \mathbf r,  \mathbf r')$. 

In the relativistic case the superconducting order parameters can be represented by $4\times4$ matrices entering the Dirac-BdG Hamiltonian,
A Lorentz symmetry analysis shows that the order parameter that transforms as a scalar under Lorentz transformations represents the relativistic generalization of BCS-type spin-singlet superconductivity, while the Balian-Werthamer parametrization of spin-triplet pairing corresponds to the axial four-vector whose spatial components reduce in the nonrelativistic limit to the Pauli spin-triplet structure $i \sigma_y (\boldsymbol{\sigma} \cdot \mathbf{d})$ on the large components of the Dirac spinor.
These considerations leads to the following generalization of the pairing potential written in $\mathbf{k}$-space
\begin{equation}
     \Delta(\mathbf{k}) =
    \underbrace{
    i \sigma_y \otimes \mathbb I_2 \, d_0(\mathbf{k})
    }_{\text{spin-singlet}}
    + \underbrace{i \sigma_y \boldsymbol{\sigma}\cdot \mathbf{d}(\mathbf{k}) \otimes \sigma_z}   
       _{\text{spin-triplet (BW)}}.   
\end{equation}

In what follows, we shall show how these equations can be solved within the
Korringa-Kohn-Rostoker band structure method
assuming the same pairing potentials corresponding to the $E_g$ and $A_{1u}$ irreps that was used for the model calculation.

\subsection{Green's function method for solving the KSDBdG equations}

To be able to treat arbitrary geometries including semi-infinite geometries without the use of a large super cell, we use a Green's function based approach exploiting the real-space representation of the resolvent of the KSDBdG Hamiltonian, 
\begin{equation}
 \mathcal{G}(z)=\left( z \mathbb{I}
 -H_{\text{DBdG}}\right)^{-1}.
 \label{eq:res}
\end{equation}

The Multiple Scattering Theory (MST), i.e. the Korringa-Kohn-Rostoker (KKR) method gives direct access to the Green's function avoiding the step of calculating Kohn-Sham orbitals. In Ref.~\cite{Csire2018} this technique was generalised for the solution of the KSDBdG equation with layered structure.
In MST, the potential is treated in the so called muffin-tin approximation, i.e. 
the potential is written as a sum of single-domain potentials centered around each lattice site, $n$, namely $V_{\text{eff}}( \mathbf r)= \sum_n V_n(r)$, 
and for the pairing potential we adopt a local approximation,
$\Delta(\mathbf r, \mathbf r') \approx \Delta_{\rm eff}(\mathbf r) \, \delta(\mathbf r-\mathbf r')$ which is treated as $\Delta_{\text{eff}}( \mathbf r)= \sum_n \Delta_{n}(r)$. The potentials treated within the atomic sphere approximation (ASA) are zero if $r=| \mathbf r_n|\geq S_n$, where $S_n$ is the radius of the Wigner-Seitz (WS) sphere that has the same volume as the atomic cell $n$.

In the relativistic case, we search the solutions of the KSDBdG equations in the following form
\begin{equation}
 \Psi(z,  \mathbf r)= \sum_{Q}
 \begin{pmatrix}
  g^e_{Q}(z,r) \chi_{Q} (\hat r) \\
  i f^e_{Q}(z,r)\chi_{\overline Q} (\hat r) \\
  g^h_{Q}(z,r) \chi^*_{Q} (\hat r) \\
  -i f^h_{Q}(z,r)\chi^*_{\overline Q} (\hat r)
 \end{pmatrix},
\end{equation}
where $Q=(\kappa,\mu)$ and $\overline{Q}=(-\kappa,\mu)$ are the composite indices for the spin-orbit ($\kappa$) and magnetic ($\mu$) quantum numbers; $g^{e(h)}_{Q}(z,r)$ and $f^{e(h)}_{Q}(z,r)$ are the large and small components of the electron (hole) part of the solution, respectively. The spin-angular function is an eigenfunction of the spin-orbit operator $ K = \boldsymbol \sigma L +\mathbb I$,
\begin{equation}
 K \ket{\kappa \mu} = -\kappa \ket{\kappa \mu}.
\end{equation}
This basis set has the advantage that the corresponding matrix of the spin-orbit operator is diagonal and it also allows an artificial scaling of spin-orbit coupling for testing relativistic effects. For later purposes the following notations are also introduced: $\overline{l}=l-S_k$, and $S_k=\kappa/|\kappa|$ the sign of $\kappa$.

With integration over the angular parts and using the orthonormality
of the Clebsch-Gordan coefficients
the radial KSDBdG equations for the right-hand side solutions within the muffin-tin spheres can be written as
\begin{equation}
H_{\textrm{radial}} =
\begin{pmatrix}
E_F & c \left( \frac{\dd}{\dd r} + \frac{1}{r} - \frac{\kappa}{r} \right) \\
c \left( \frac{\dd}{\dd r} + \frac{1}{r} + \frac{\kappa}{r} \right) & E_F + c^2
\end{pmatrix}
\end{equation}
\begin{equation}
\begin{split}
&
\begin{pmatrix}
H_{\textrm{radial}} + z \mathbb I_2 & 0 \\
0 & - H_{\textrm{radial}}  + z \mathbb I_2
\end{pmatrix}
\begin{pmatrix}
g^e_{Q}(z,r) \\
f^e_{Q}(z,r) \\
g^h_{Q}(z,r) \\
f^h_{Q}(z,r)
\end{pmatrix}
 = \\
&
\sum_{Q'}
\begin{pmatrix}
0 & \Delta_{Q Q'}(r) \mathbb I_2   \\
\Delta^*_{Q' Q}(r) \mathbb I_2 & 0 
\end{pmatrix}
\begin{pmatrix}
g^e_{Q'}(z,r) \\
f^e_{Q'}(z,r) \\
g^h_{Q'}(z,r) \\
f^h_{Q'}(z,r)
\end{pmatrix},
\end{split}
\end{equation}
with
\begin{equation}
\Delta_{QQ'} = \sum_{s,s' = \pm 1/2} 
c(\ell j \tfrac{1}{2}; \mu - s, s)\,
\Delta^{ss'}_{LL'}\,
c(\ell' j' \tfrac{1}{2}; \mu' - s', s'),
\end{equation}
where $ L = (\ell, \mu - s), L' = (\ell', \mu' - s')$.
Thus, in general, we can implement orbital-dependent pairing of arbitrary type
which allows us to realize the same pairing interactions as introduced in the S1 part.

The two most important quantities leading to the Green's function are the single-site $t$-matrix and the structure constants. Physically, the $t$-matrix describes scattering on the the single-site potential, while the structure constants contain all information about the crystal structure. We shall denote the irregular and regular solutions of the KSDBdG equations inside the WS spheres by $\mathbf{J}_Q(z,\mathbf{r})$ and $\mathbf{Z}_Q(z,\mathbf{r})$ (matrices in electron-hole indices), respectively. The scattering solutions obey the following matching conditions (derived from the Lippmann-Schwinger equations as described in Ref.~\onlinecite{Csire2018}.

\begin{subequations}
\begin{eqnarray}
    J_Q^{ab}(z,\mathbf{r}) &=& j_Q^a (z,\mathbf{r}) \delta_{ab},\\
    Z_Q^{ab}(z,\mathbf{r})&=& \sum_{Q'} j_{Q'}^{a} (z,\mathbf{r}) m^{ab}_{Q'Q}(z)
    -i p^a h_{Q}^{a}(z,\mathbf{r})\delta_{ab},\qquad
  \end{eqnarray}%
  \label{eq:GF_calcrel}%
\end{subequations}%
where $j_Q^a (z,\mathbf{r})$ and $h_{Q}^{a}(z,\mathbf{r})$ are spherical Bessel and Hankel type of solutions for the case of zero potential, respectively, and we introduced the inverse $t$-matrix, 
$m_{QQ'}^{ab}(z)=\left[\underline{t}^{-1}(z)\right]_{QQ'}^{ab}$.

As it was described in Ref.~\onlinecite{Csire2018}, the relativistic real space structure constants should be derived from their non-relativistic counterpart $G_{0,LL'}^{\textrm{NR},ee,nm}(z) \delta_{ss'}$ with $L=(\ell,m)$ and $s$ being angular momentum and spin indices, respectively, by transforming it into the $\ket{\kappa \mu}$ basis, while the hole part of the relativistic structure constants can be constructed in the $\ket{\kappa \mu}^*$ basis using the non-relativistic formula\cite{Csire2015} $G_{0,LL'}^{\textrm{NR},hh,nm}(z)=-G_{0,LL'}^{\textrm{NR},ee,nm}(-z)$.

The following supermatrix formalism can be introduced for the scattering matrices, the matrices of the structure constant and the scattering path operator:
\begin{eqnarray}
 \mathbf{t}(z) &=& \{t^{n,ab}_{QQ'}(z) \delta_{nm} \},
 \label{eq:sh_t}\\
 \mathbf{G}_0(z) &=& \{G^{nm,ab}_{0,QQ'}(z) (1-\delta_{nm}) \delta_{ab} \},
 \label{eq:sh_G_0}\\
 \boldsymbol{\tau}(z) &=& \{\tau^{nm,ab}_{QQ'}(z) \},
 \label{eq:sh_tau}
\end{eqnarray}
where $\boldsymbol{\tau}(z)$ can be determined from the single-site t-matrix and the real space structure constant matrix 
\begin{equation}
 \boldsymbol{\tau}(z)=
 \left(
 \mathbf{t}(z)^{-1} -\mathbf{G}_0(z)
 \right)^{-1} \, .
  \label{eq:sh_tau_t}
\end{equation}

As in the normal state formalism, based on the expansions of the free-particle Green's function, 
the derivation of the one-particle Green's function  is straightforward and leads to the following formula,
\begin{equation}
\begin{split}
 \mathbf{G}(z ,\mathbf r,\mathbf {r'}) &=
   \sum_{QQ'} \mathbf{Z}_Q(z ,\mathbf r) \boldsymbol{\tau}_{QQ'}(z )
   \mathbf{Z}_{Q'}(z ,\mathbf{r'})^\times
 \\ &- \theta(r'- r) \sum_Q \mathbf{Z}_Q(z ,\mathbf r) \mathbf{J}_Q(z ,\mathbf{r'})^\times
 \\ &- \theta(r - r') \sum_Q \mathbf{J}_Q(z ,\mathbf r) \mathbf{Z}_Q(z ,\mathbf{r'})^\times,
\end{split}
  \label{eq:gf_formula}
\end{equation}
where $\theta(x)$ is the step function and
\begin{equation}
\begin{split}
\mathbf{Z}_Q(z ,\mathbf{r'})^\times
&= \mathbf{Z}_Q(z^*,\mathbf{r'})^\dagger \\
\mathbf{J}_Q(z ,\mathbf{r'})^\times
&= \mathbf{J}_Q(z^*,\mathbf{r'})^\dagger \, .
\end{split}
  \label{eq:Ztimes}
\end{equation}
stand for the left-hand side solutions of the KSDBdG equations assuming that the pairing potential is not complex.
However, as we implement the chiral $d$-wave ($E_g$) pairing which breaks time-reversal symmetry and non-trivially complex we also need to calculate directly the left-hand side solutions
$\mathbf{Z}_Q(z ,\mathbf{r'})^\times$ and $\mathbf{J}_Q(z ,\mathbf{r'})^\times$ solutions
in the same way as was described in Ref.~\onlinecite{Csire2018}.

The formulas given above can be applied to surfaces and interfaces following the so-called Screened KKR (SKKR) formalism described in Refs.~\onlinecite{Szunyogh, Zeller}. In this formalism, it is made use of the 2D periodicity of the layers by introducing 2D lattice Fourier transformed version for the scattering path operator
\begin{equation}
 \boldsymbol{\tau}(z,\mathbf k_{||})=
 \left(
 \mathbf{t}(z)^{-1} -\mathbf{G}_0(z,\mathbf k_{||})
 \right)^{-1}.
  \label{eq:sh_tau_t2}
\end{equation}
To perform the inverse of the KKR matrix, a special reference system is used to obtain structure constants that are localized in real space, hence, the formalism can be derived for layered systems with two-dimensional periodicity and applied as the SKKR method prescribes\cite{Szunyogh, Zeller} yielding the Green's function for surfaces/interfaces.
Then the DOS can be calculated as
\begin{equation}
D(\varepsilon) = -\frac{1}{\pi \, \Omega_{\mathrm{BZ}}} 
\int_{\mathrm{BZ}} d\mathbf{k}_\parallel \int d^3 \mathbf{r} \; 
\mathrm{Im} \, \mathrm{Tr} \, G(\varepsilon +i0, \mathbf{k}_\parallel, \mathbf{r}, \mathbf{r}).
\end{equation}

\subsection{Modeling details}

Sr$_2$RuO$_4$ is a tetragonal layered perovskite with space group $I4/mmm$. The lattice constants of the layered perovskite strontium ruthenate are $a = 3.87$~\AA{} and $c = 12.74$~\AA{}.
First, normal state, conventional DFT calculations are performed with the SKKR Green's function formalism based on the local-density approximation (LDA).
We begin with a bulk calculation to obtain self-consistent electrostatic potentials and the Fermi energy for the bulk Sr$_2$RuO$_4$. Then, normal state surface calculations are done to obtain potentials and vacuum potential level for the different surfaces of the semi-infinite host and also using Ag overlayers.
Silver has a face-centered cubic (fcc) crystal structure with a lattice constant of $a_{\mathrm{Ag}} = 4.08$~\AA, 
which corresponds to approximately $5\%$ compressive strain when an Ag(001) overlayer is matched epitaxially to the in-plane lattice constant of Sr$_2$RuO$_4$. 
This lattice matching provides a simple model for a coherent Ag/Sr$_2$RuO$_4$ interface. 
Two registry configurations of the Ag(001) overlayer are considered, corresponding to the two possible ways the fcc stacking can sit on the Sr$_2$RuO$_4$ surface: the first Ag layer is either aligned with surface O atoms or with surface Ru atoms. 
In the latter case, the fcc stacking places an additional Ag atom in the center of the square formed by the first layer, modifying the local environment and hybridization at the interface.

After having the normal state potentials for the surfaces and interfaces, the Kohn-Sham-Dirac-Bogoliubov-de Gennes are solved assuming the interobital $E_g$ and $A_{1u}$ pairings within the same KKR Green's function formalism.
In our calculations, we assume a superconducting gap of $\Delta = 0.5$ meV, which is slightly larger than the typical experimental values ($\sim 0.25$-$0.35$ meV).
This choice reduces the computational demand and enhances the visibility of quasiparticle features in the calculated spectra, such as zero-energy and coherence peaks.
Importantly, the qualitative behavior of the spectra -- including broadening, surface dependence, and hybridization effects -- remain unaffected by this modest increase in the superconducting gap.

\section{S3: Formulation of surface density of states in the three-dimensional $\mathrm{Sr}_2\mathrm{RuO}_4$ model}

In this section, we provide details of the model calculation about the surface and planar edge density of states in Sr$_2$RuO$_4$ in the main text.
We deal with the three-dimensional (3D) Sr$_2$RuO$_4$ model suggested in Ref.\, \cite{SuhPRR2020}.
In the bulk, the normal state Hamiltonian is given by
\begin{align}
    \hat{H}(\bm{k})&=\sum^{8}_{i=0}\sum^{3}_{j=0}h_{i,j}(\bm{k})\hat{L}_{i}\otimes\hat{\sigma}_{j},
\end{align}%
where $\hat{L}_{i}$ are the Gell-Mann matrices described by
\begin{align}
    \hat{L}_{0}&=
    \begin{pmatrix}
        1 & 0 & 0 \\
        0 & 1 & 0 \\
        0 & 0 &1
    \end{pmatrix},
\end{align}%
\begin{align}
    \hat{L}_{1}&=
    \begin{pmatrix}
        0 & 1 & 0 \\
        1 & 0 & 0 \\
        0 & 0 &0
    \end{pmatrix},
\end{align}%
\begin{align}
    \hat{L}_{2}&=
    \begin{pmatrix}
        0 & 0 & 1 \\
        0 & 0 & 0 \\
        1 & 0 & 0
    \end{pmatrix},
\end{align}%
\begin{align}
    \hat{L}_{3}&=
    \begin{pmatrix}
        0 & 0 & 0 \\
        0 & 0 & 1 \\
        0 & 1 & 0
    \end{pmatrix},
\end{align}%
\begin{align}
    \hat{L}_{4}&=
    \begin{pmatrix}
        0 & -i & 0 \\
        i & 0 & 0 \\
        0 & 0 &0
    \end{pmatrix},
\end{align}%
\begin{align}
    \hat{L}_{5}&=
    \begin{pmatrix}
        0 & 0 & -i \\
        0 & 0 & 0 \\
        i & 0 & 0
    \end{pmatrix},
\end{align}%
\begin{align}
    \hat{L}_{6}&=
    \begin{pmatrix}
        0 & 0 & 0 \\
        0 & 0 & -i \\
        0 & i & 0
    \end{pmatrix},
\end{align}%
\begin{align}
    \hat{L}_{7}&=
    \begin{pmatrix}
        1 & 0 & 0 \\
        0 & -1 & 0 \\
        0 & 0 &0
    \end{pmatrix},
\end{align}%
\begin{align}
    \hat{L}_{8}&=\frac{1}{\sqrt{3}}
    \begin{pmatrix}
        1 & 0 & 0 \\
        0 & 1 & 0 \\
        0 & 0 & -2
    \end{pmatrix},
\end{align}%
in the basis $[d_{yz},d_{zx},d_{xy}]$-orbitals, and $\hat{\sigma}_{j}$ is the Pauli matrices in spin space.
$h_{i,j}$ are described by
\begin{align}
    h_{0,0}(\bm{k})=\frac{1}{3}[\xi_{11}(\bm{k})+\xi_{22}(\bm{k})+\xi_{33}(\bm{k})],
\end{align}%
\begin{align}
    h_{8,0}(\bm{k})=\frac{1}{2\sqrt{3}}[\xi_{11}(\bm{k})+\xi_{22}(\bm{k})-2\xi_{33}(\bm{k})],
\end{align}%
\begin{align}
    h_{7,0}(\bm{k})=\frac{1}{2}[\xi_{22}(\bm{k})-\xi_{11}(\bm{k})],
\end{align}%
for the intraorbital hopping, 
\begin{align}
    h_{1,0}(\bm{k})=\lambda_{z}(\bm{k}),
\end{align}%
\begin{align}
    h_{2,0}(\bm{k})=8t^{z}_{13}\sin\frac
    {k_z}{2}\sin\frac{k_x}{2}\cos\frac{k_y}{2},
\end{align}%
\begin{align}
    h_{3,0}(\bm{k})=8t^{z}_{13}\sin\frac
    {k_z}{2}\cos\frac{k_x}{2}\sin\frac{k_y}{2},
\end{align}%
for the interorbital hopping,
\begin{align}
    h_{5,2}(\bm{k})=\eta_\mathrm{so}+2\gamma^\mathrm{so}_{5261}[\cos{k_x}-\cos{k_y}],
\end{align}%
\begin{align}
    h_{6,1}(\bm{k})=-\eta_\mathrm{so}+2\gamma^\mathrm{so}_{5261}[\cos{k_x}-\cos{k_y}],
\end{align}%
\begin{align}
    h_{4,3}(\bm{k})=-\eta_\mathrm{so},
\end{align}%
\begin{align}
    h_{5,1}(\bm{k})=-h_{6,2}(\bm{k})=4\gamma^\mathrm{so}_{5162}\sin{k_x}\sin{k_y},
\end{align}%
\begin{align}
    h_{4,1}(\bm{k})=8\gamma^\mathrm{so}_{12,z}\sin\frac
    {k_z}{2}\sin\frac{k_x}{2}\cos\frac{k_y}{2},
\end{align}%
\begin{align}
    h_{4,2}(\bm{k})=8\gamma^\mathrm{so}_{12,z}\sin\frac
    {k_z}{2}\cos\frac{k_x}{2}\sin\frac{k_y}{2},
\end{align}%
\begin{align}
    h_{6,3}(\bm{k})=8\gamma^\mathrm{so}_{56,z}\sin\frac
    {k_z}{2}\sin\frac{k_x}{2}\cos\frac{k_y}{2},
\end{align}%
\begin{align}
    h_{5,3}(\bm{k})=-8\gamma^\mathrm{so}_{56,z}\sin\frac
    {k_z}{2}\cos\frac{k_x}{2}\sin\frac{k_y}{2},
\end{align}%
for the atomic and momentum-dependent spin-orbit coupling terms.
Here, $\xi_{11}(\bm{k})$, $\xi_{22}(\bm{k})$, $\xi_{33}(\bm{k})$ with $2=d_{yz}$, $1=d_{zx}$, and $3=d_{xy}$, and $\lambda_{z}(\bm{k})$ are
\begin{align}
    \xi_{11}(\bm{k})&=-\mu+2t_{11}^{x}\cos{k_x}+2t_{11}^{y}\cos{k_y}\notag\\
    &+8t^{z}_{11}\cos\frac{k_z}{2}\cos\frac{k_x}{2}\cos\frac{k_y}{2}\notag\\
    &+4t^{xy}_{11}\cos{k_x}\cos{k_y}\notag\\
    &+2t^{xx}_{11}\cos{2k_x}+2t^{yy}_{11}\cos{2k_y}\notag\\
    &+4t^{xxy}_{11}\cos{2k_x}\cos{k_y}\notag\\
    &+4t^{xyy}_{11}\cos{k_x}\cos{2k_y}\notag\\
    &+2t^{zz}_{11}[\cos{k_z}-1],
\end{align}%
\begin{align}
    \xi_{22}(\bm{k})&=-\mu+2t_{11}^{y}\cos{k_x}+2t_{11}^{x}\cos{k_y}\notag\\
    &+8t^{z}_{11}\cos\frac{k_z}{2}\cos\frac{k_x}{2}\cos\frac{k_y}{2}\notag\\
    &+4t^{xy}_{11}\cos{k_x}\cos{k_y}\notag\\
    &+2t^{yy}_{11}\cos{2k_x}+2\gamma^{xx}_{11}\cos{2k_y}\notag\\
    &+4t^{xyy}_{11}\cos{2k_x}\cos{k_y}\notag\\
    &+4t^{xxy}_{11}\cos{k_x}\cos{2k_y}\notag\\
    &+2t^{zz}_{11}[\cos{k_z}-1],
\end{align}%
\begin{align}
    \xi_{33}(\bm{k})&=-\mu_{3}+2t_{33}^{x}(\cos{k_x}+\cos{k_y})\notag\\
    &+8t^{z}_{33}\cos\frac{k_z}{2}\cos\frac{k_x}{2}\cos\frac{k_y}{2}\notag\\
    &+4t^{xy}_{33}\cos{k_x}\cos{k_y}\notag\\
    &+2t^{xx}_{33}(\cos{2k_x}+\cos{2k_y})\notag\\
    &+4t^{xxy}_{33}(\cos{2k_x}\cos{k_y}+\cos{k_x}\cos{2k_y})\notag\\
    &+2t^{zz}_{33}[\cos{k_z}-1],
\end{align}%
\begin{align}
    \lambda_{z}(\bm{k})&=4t^{z}_{12}\cos\frac{k_z}{2}\sin\frac{k_x}{2}\sin\frac{k_y}{2}\notag\\
    &-4t^{xy}_{12}\sin{k_x}\sin{k_y}\notag\\
    &-4t^{xxy}_{12}(\sin{2k_x}\sin{k_y}+\sin{k_x}\sin{2k_y}).
\end{align}%
The parameters are in unit of meV  \cite{SuhPRR2020}): $t^{x}_{11}=-362.4$, $t^{y}_{11}=-134$, $t^{x}_{33}=-262.4$, $t^{xy}_{11}=-44.01$, $t^{xx}_{11}=-1.021$, $t^{yy}_{11}=-5.727$, $t^{xy}_{33}=-43.73$, $t^{xx}_{33}=34.23$, $t^{xxy}_{11}=-13.93$, $t^{xyy}_{11}=-7.52$, $t^{xxy}_{33}=8.069$, $t^{z}_{11}=-0.0228$, $t^{z}_{33}=1.811$, $t^{z}_{13}=8.304$, $t^{zz}_{11}=2.522$, $t^{zz}_{33}=-3.159$, $\mu=438.5$, $\mu_{3}=218.6$, $t^{z}_{12}=19.95$, $t^{xy}_{12}=16.25$, $t^{xxy}_{12}=3.94$, $\eta_\mathrm{so}=57.39$, $\gamma^\mathrm{so}_{56,z}=-1.247$, $\gamma^\mathrm{so}_{12,z}=-3.576$, $\gamma^\mathrm{so}_{5162}=-1.008$, and $\gamma^\mathrm{so}_{5261}=0.3779$.

In the superconducting state, we consider interorbital pairing symmetries in the above model.
Then the Bogoliubov-de Gennes Hamiltonian is given by
\begin{align}
    \hat{H}_\mathrm{BdG}(\bm{k})=
    \begin{pmatrix}
        \hat{H}(\bm{k}) & \hat{\Delta}(\bm{k})\\
        \hat{\Delta}^{\dagger}(\bm{k}) & -\hat{H}^{*}(-\bm{k})
    \end{pmatrix},
\end{align}
with the pair potential $\hat{\Delta}(\bm{k})$.
Based on Refs.\ \cite{SuhPRR2020,FukayaPRR02022}, we choose $\Delta(\bm{k})$ as:
\begin{align}
    \hat{\Delta}^\mathrm{T}=\Delta[(-\hat{L}_{6}+i\hat{L}_{5})\otimes\hat{\sigma}_{3}]i\hat{\sigma}_{2},
\end{align}%
for the interorbital spin-triplet $E_g$~\cite{SuhPRR2020}
and 
\begin{align}
    \hat{\Delta}^\mathrm{S}(\bm{k})=-\Delta[\hat{L}_{4}\otimes\hat{\sigma}_{0}\sin{k_z}]i\hat{\sigma}_{2},
\end{align}%
for the interorbital spin-singlet A$_{1u}$ pairings~\cite{FukayaPRR02022}.

\begin{figure}[t!]
    \centering
    \includegraphics[width=8.5cm]{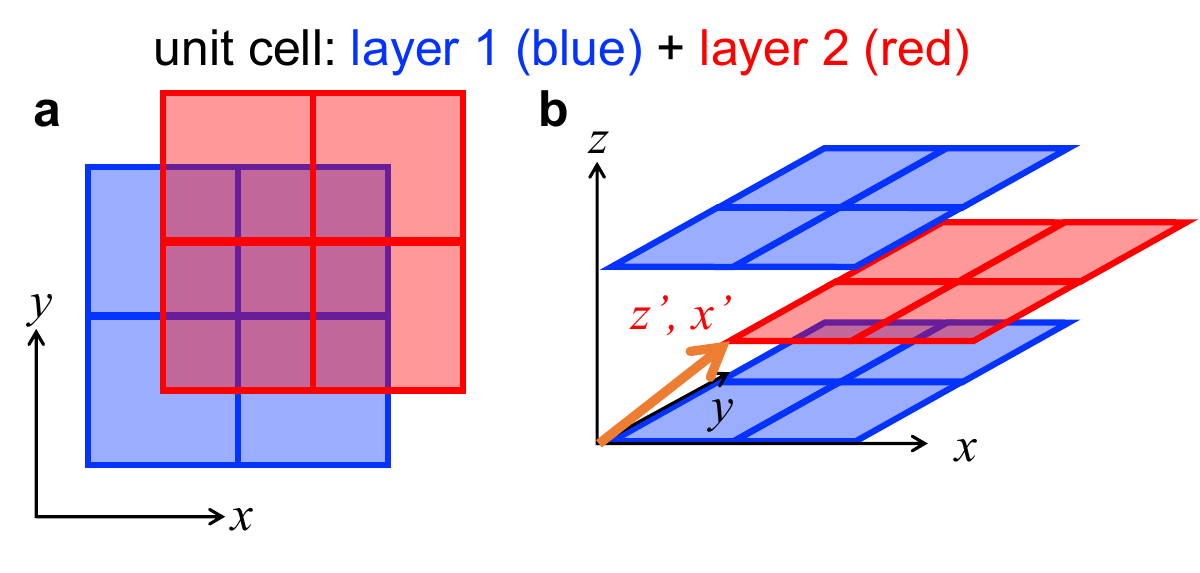}
    \caption{\textbf{Image of transformation in the real space.} (a) View from the $z$-direction. Layer 2 with red color misaligns for $(x',y',z')=(a/2,a/2,c/2)$ with the lattice constants $a=c=1$.}
    \label{fig:lattice_t}
\end{figure}%

We assume the surface along the $z$ and $x$-directions and the semi-infinite superconductors.
We deal with the recursive Green's function method~\cite{Umerski} to calculate the surface density of states (SDOS) for each direction.
In the real space, the 3D Sr$_2$RuO$_4$ model in Ref.\ \cite{SuhPRR2020} is composed of two layers as shown in Fig.~\ref{fig:lattice_t} (a).
Then the coordinate of layer 2 misaligns for $(x',y',z')=(a/2,a/2,c/2)$ with the lattice constants set by $a=c=1$, and the second (fourth) nearest-neighbor hopping is included for the Fourier transformation along the $z$ ($x$)-direction.
For this reason, we consider the transformation:
\begin{align}
    k_{z',x'}=\frac{k_x}{2}+\frac{k_y}{2}+\frac{k_z}{2},
\end{align}%
along the $z$ and the $x$-direction shown in Fig.~\ref{fig:lattice_t} (b).
For the Fourier transformation along the $z'$ and $x'$-directions, we can calculate the surface Green's function by using the recursive Green's function method~\cite{Umerski,Takagi20}.
We note that this transformation is topologically the same as at the surface.

\section{S4: Pair potential analysis in the band basis}

In this section, we demonstrate how the the pair potential is modified when considering its structure in the band basis.
For the interorbital E$_g$ pairing $\{[5,3],[6,3]\}$, we assume the following normal state Hamiltonian:
\begin{align}
    \hat{H}^{1}_{eff}(\bm{k})&=h_{5,3}(\bm{k})\hat{L}_{5}\otimes\hat{\sigma}_{3}+h_{6,3}(\bm{k})\hat{L}_{6}\otimes\hat{\sigma}_{3}.
\end{align}%
Then we obtain the unitary operator composed with the eigenstates:
\begin{align}
    \hat{U}_{1}(\bm{k})=
    \begin{pmatrix}
    \hat{U}^{1}_{\uparrow\uparrow}(\bm{k}) & 0 \\
    0 &\hat{U}^{1}_{\downarrow\downarrow}(\bm{k})
    \end{pmatrix},
\end{align}%
with
\begin{align}
    \hat{U}^{1}_{\uparrow\uparrow}(\bm{k})=
\begin{pmatrix}
-i\frac{h_{6,3}}{\sqrt{2(h_{6,3}^{2}+h^{2}_{5,3})}} & \frac{h_{5,3}}{\sqrt{h_{6,3}^{2}+h^{2}_{5,3}}} & i\frac{h_{6,3}}{\sqrt{2(h_{6,3}^{2}+h^{2}_{5,3})}}\\
i\frac{h_{5,3}}{\sqrt{2(h_{6,3}^{2}+h^{2}_{5,3})}} & \frac{h_{6,3}}{\sqrt{h_{6,3}^{2}+h^{2}_{5,3}}} & -i\frac{h_{5,3}}{\sqrt{2(h_{6,3}^{2}+h^{2}_{5,3})}} \\
1/\sqrt{2} & 0 & 1/\sqrt{2}
\end{pmatrix},
\end{align}%
and
\begin{align}
    \hat{U}^{1}_{\downarrow\downarrow}(\bm{k})=
\begin{pmatrix}
i\frac{h_{6,3}}{\sqrt{2(h_{6,3}^{2}+h^{2}_{5,3})}} & \frac{h_{5,3}}{\sqrt{h_{6,3}^{2}+h^{2}_{5,3}}} & -i\frac{h_{6,3}}{\sqrt{2(h_{6,3}^{2}+h^{2}_{5,3})}}\\
-i\frac{h_{5,3}}{\sqrt{2(h_{6,3}^{2}+h^{2}_{5,3})}} & \frac{h_{6,3}}{\sqrt{h_{6,3}^{2}+h^{2}_{5,3}}} & i\frac{h_{5,3}}{\sqrt{2(h_{6,3}^{2}+h^{2}_{5,3})}} \\
1/\sqrt{2} & 0 & 1/\sqrt{2}
\end{pmatrix}.
\end{align}%
By this unitary operator, we also obtain 
\begin{align}
    \hat{U}^{\dagger}_{1}(\bm{k})\hat{H}^{1}_{eff}(\bm{k})\hat{U}_{1}(\bm{k})=
    \begin{pmatrix}
    \eta_{1}(\bm{k}) & 0 & 0 \\
    0 & 0 & 0 \\
    0 & 0 & -\eta_{1}(\bm{k})
\end{pmatrix}
\otimes\hat{\sigma}_{0},
\end{align}%
with 
\begin{align}
    \eta_{1}(\bm{k})=-\sqrt{h^{2}_{6,3}+h^{2}_{5,3}},
\end{align}%
and
\begin{align}
    \hat{U}^{\dagger}_{1}(\bm{k})[\Delta\{-\hat{L}_{6}\otimes\hat{\sigma}_{3}+i\hat{L}_{5}\otimes\hat{\sigma}_{3}\}]\hat{U}_{1}(\bm{k})\rightarrow\tilde{\Delta}_{1}(\bm{k}),
\end{align}%
with $\hat{\Delta}_{1}=[-\hat{L}_{6}\otimes\hat{\sigma}_{3}+i\hat{L}_{5}\otimes\hat{\sigma}_{3}]i\hat{\sigma}_{2}$ and
\begin{align}
    \tilde{\Delta}_{1}(\bm{k})=
\begin{pmatrix}
\tilde{\Delta}^{\uparrow\uparrow}_{1}(\bm{k}) & 0\\
0 & \tilde{\Delta}^{\downarrow\downarrow}_{1}(\bm{k})
\end{pmatrix},
\end{align}%
in the band basis.
Indeed, we obtain:
\begin{align}
    \tilde{\Delta}^{\uparrow\uparrow}_{1}(\bm{k})=&& \\&& \hspace{-2cm}
    \begin{pmatrix}
    -\Delta\frac{h_{6,3}+ih_{5,3}}{\sqrt{h_{6,3}^{2}+h^{2}_{5,3}}} & -\Delta\frac{h_{6,3}+ih_{5,3}}{\sqrt{2(h_{6,3}^{2}+h^{2}_{5,3})}} &0 \\
    \Delta\frac{h_{6,3}+ih_{5,3}}{\sqrt{2(h_{6,3}^{2}+h^{2}_{5,3})}} & 0 & \Delta\frac{h_{6,3}+ih_{5,3}}{\sqrt{2(h_{6,3}^{2}+h^{2}_{5,3})}}\\
    0 & -\Delta\frac{h_{6,3}+ih_{5,3}}{\sqrt{2(h_{6,3}^{2}+h^{2}_{5,3})}} & \Delta\frac{h_{6,3}+ih_{5,3}}{\sqrt{h_{6,3}^{2}+h^{2}_{5,3}}}
\end{pmatrix},
\end{align}%
and
\begin{align}
    \tilde{\Delta}^{\downarrow\downarrow}_{1}(\bm{k})=&& \\&& \hspace{-2cm}
    \begin{pmatrix}
    -\Delta\frac{h_{6,3}+ih_{5,3}}{\sqrt{h_{6,3}^{2}+h^{2}_{5,3}}} & \Delta\frac{h_{6,3}+ih_{5,3}}{\sqrt{2(h_{6,3}^{2}+h^{2}_{5,3})}} &0 \\
    -\Delta\frac{h_{6,3}+ih_{5,3}}{\sqrt{2(h_{6,3}^{2}+h^{2}_{5,3})}} & 0 & -\Delta\frac{h_{6,3}+ih_{5,3}}{\sqrt{2(h_{6,3}^{2}+h^{2}_{5,3})}}\\
    0 & \Delta\frac{h_{6,3}+ih_{5,3}}{\sqrt{2(h_{6,3}^{2}+h^{2}_{5,3})}} & \Delta\frac{h_{6,3}+ih_{5,3}}{\sqrt{h_{6,3}^{2}+h^{2}_{5,3}}}
\end{pmatrix}.
\end{align}%

While, for the interorbital spin-triplet B$_{1g}$ pairing with $\hat{\Delta}_{B_{1g}}=[\hat{L}_{5}\otimes\hat{\sigma}_{2}+i\hat{L}_{6}\otimes\hat{\sigma}_{1}]i\hat{\sigma}_{2}$, we obtain:
\begin{align}
    \tilde{\Delta}_{B_{1g}}(\bm{k})=
\begin{pmatrix}
0 &\tilde{\Delta}^{\uparrow\downarrow}_{B_{1g}}(\bm{k})\\
\tilde{\Delta}^{\downarrow\uparrow}_{B_{1g}}(\bm{k}) & 0
\end{pmatrix},
\end{align}%
with
\begin{align}
    \tilde{\Delta}^{\uparrow\downarrow}_{B_{1g}}(\bm{k})=&& \\&& \hspace{-2cm}
    \begin{pmatrix}
    0 & \Delta\frac{h_{6,3}-ih_{5,3}}{\sqrt{2(h_{6,3}^{2}+h^{2}_{5,3})}} &i\Delta\frac{h_{6,3}-ih_{5,3}}{\sqrt{h_{6,3}^{2}+h^{2}_{5,3}}} \\
    \Delta\frac{-h_{6,3}+ih_{5,3}}{\sqrt{2(h_{6,3}^{2}+h^{2}_{5,3})}} & 0 & \Delta\frac{-h_{6,3}+ih_{5,3}}{\sqrt{2(h_{6,3}^{2}+h^{2}_{5,3})}}\\
    -i\Delta\frac{h_{6,3}-ih_{5,3}}{\sqrt{h_{6,3}^{2}+h^{2}_{5,3}}} & \Delta\frac{h_{6,3}-ih_{5,3}}{\sqrt{2(h_{6,3}^{2}+h^{2}_{5,3})}} & 0
\end{pmatrix},
\end{align}%
and
\begin{align}
    &\tilde{\Delta}^{\downarrow\uparrow}_{B_{1g}}(\bm{k})\\
    &=
    \begin{pmatrix}
    0 & -\Delta\frac{h_{6,3}+ih_{5,3}}{\sqrt{2(h_{6,3}^{2}+h^{2}_{5,3})}} &-i\Delta\frac{h_{6,3}+ih_{5,3}}{\sqrt{h_{6,3}^{2}+h^{2}_{5,3}}} \\
    \Delta\frac{h_{6,3}+ih_{5,3}}{\sqrt{2(h_{6,3}^{2}+h^{2}_{5,3})}} & 0 & \Delta\frac{h_{6,3}-ih_{5,3}}{\sqrt{2(h_{6,3}^{2}+h^{2}_{5,3})}}\\
    i\Delta\frac{h_{6,3}+ih_{5,3}}{\sqrt{h_{6,3}^{2}+h^{2}_{5,3}}} & \Delta\frac{-h_{6,3}+ih_{5,3}}{\sqrt{2(h_{6,3}^{2}+h^{2}_{5,3})}} & 0
\end{pmatrix}.\notag
\end{align}%
This result indicates that the E$_{g}$ channel in the normal state does not contribute to the interorbital B$_{1g}$ state as the intraband pairings.

As well as for another E$_g$ pairing, we obtain the pair potential in the band basis by focusing on the $t^{z}_{13}$ term.
We consider the normal state Hamiltonian at the Fermi level:
\begin{align}
    \hat{H}^{2}_{eff}(\bm{k})&=h_{2,0}(\bm{k})\hat{L}_{2}\otimes\hat{\sigma}_{0}+h_{3,0}(\bm{k})\hat{L}_{3}\otimes\hat{\sigma}_{0}.
\end{align}%
The corresponding unitary operator is given by
\begin{align}
    \hat{U}_{2}(\bm{k})=
    \begin{pmatrix}
    \hat{U}^{2}_{\uparrow\uparrow}(\bm{k}) & 0 \\
    0 &\hat{U}^{2}_{\downarrow\downarrow}(\bm{k})
    \end{pmatrix},
\end{align}%
with
\begin{align}
    \hat{U}^{2}_{\uparrow\uparrow}(\bm{k})=
\begin{pmatrix}
-\frac{h_{3,0}}{\sqrt{2(h_{2,0}^{2}+h^{2}_{3,0})}} & -\frac{h_{2,0}}{\sqrt{h_{2,0}^{2}+h^{2}_{3,0}}} & \frac{h_{3,0}}{\sqrt{2(h_{2,0}^{2}+h^{2}_{3,0})}}\\
-\frac{h_{2,0}}{\sqrt{2(h_{2,0}^{2}+h^{2}_{3,0})}} & \frac{h_{3,0}}{\sqrt{h_{2,0}^{2}+h^{2}_{3,0}}} & \frac{h_{2,0}}{\sqrt{2(h_{2,0}^{2}+h^{2}_{3,0})}} \\
1/\sqrt{2} & 0 & 1/\sqrt{2}
\end{pmatrix},
\end{align}%
and
\begin{align}
    \hat{U}^{2}_{\downarrow\downarrow}(\bm{k})=
\begin{pmatrix}
-\frac{h_{3,0}}{\sqrt{2(h_{2,0}^{2}+h^{2}_{3,0})}} & -\frac{h_{2,0}}{\sqrt{h_{2,0}^{2}+h^{2}_{3,0}}} & \frac{h_{3,0}}{\sqrt{2(h_{2,0}^{2}+h^{2}_{3,0})}}\\
-\frac{h_{2,0}}{\sqrt{2(h_{2,0}^{2}+h^{2}_{3,0})}} & \frac{h_{3,0}}{\sqrt{h_{2,0}^{2}+h^{2}_{3,0}}} & \frac{h_{2,0}}{\sqrt{2(h_{2,0}^{2}+h^{2}_{3,0})}} \\
1/\sqrt{2} & 0 & 1/\sqrt{2}
\end{pmatrix}.
\end{align}%
By using this unitary operator, the effective normal state Hamiltonian $\hat{H}^{2}_{eff}(\bm{k})$ is diagonalized as
\begin{align}
    \hat{U}^{\dagger}_{2}(\bm{k})\hat{H}^{2}_{eff}(\bm{k})\hat{U}_{2}(\bm{k})=
    \begin{pmatrix}
    \eta_{2}(\bm{k}) & 0 & 0 \\
    0 & 0 & 0 \\
    0 & 0 & -\eta_{2}(\bm{k})
\end{pmatrix}
\otimes\hat{\sigma}_{0},
\end{align}%
with 
\begin{align}
    \eta_{2}(\bm{k})=-\sqrt{h^{2}_{2,0}+h^{2}_{3,0}}.
\end{align}%

\begin{align}
    \hat{U}^{\dagger}_{2}(\bm{k})[\Delta\{\hat{L}_{3}\otimes\hat{\sigma}_{0}+i\hat{L}_{2}\otimes\hat{\sigma}_{0}\}]\hat{U}_{2}(\bm{k})\rightarrow\tilde{\Delta}_{1}(\bm{k}),
\end{align}%
with $\hat{\Delta}_{2}=[\hat{L}_{3}\otimes\hat{\sigma}_{0}+i\hat{L}_{2}\otimes\hat{\sigma}_{0}]i\hat{\sigma}_{2}$ and
\begin{align}
    \tilde{\Delta}_{2}(\bm{k})=
\begin{pmatrix}
\tilde{\Delta}^{\uparrow\uparrow}_{2}(\bm{k}) & 0\\
0 & \tilde{\Delta}^{\downarrow\downarrow}_{2}(\bm{k})
\end{pmatrix},
\end{align}%
in the band basis.
Thus, we obtain:
\begin{align}
    &\tilde{\Delta}^{\uparrow\uparrow}_{2}(\bm{k})=\tilde{\Delta}^{\downarrow\downarrow}_{2}(\bm{k})\\
    &=
    \begin{pmatrix}
    -i\Delta\frac{h_{3,0}-ih_{2,0}}{\sqrt{h_{2,0}^{2}+h^{2}_{3,0}}} & -\Delta\frac{h_{3,0}-ih_{2,0}}{\sqrt{2(h_{6,3}^{2}+h^{2}_{5,3})}} &0 \\
    \Delta\frac{h_{3,0}-ih_{2,0}}{\sqrt{2(h_{2,0}^{2}+h^{2}_{3,0})}} & 0 & -\Delta\frac{h_{3,0}-ih_{2,0}}{\sqrt{2(h_{2,0}^{2}+h^{2}_{3,0})}}\\
    0 & -\Delta\frac{h_{3,0}-ih_{2,0}}{\sqrt{2(h_{2,0}^{2}+h^{2}_{3,0})}} & i\Delta\frac{h_{3,0}-ih_{2,0}}{\sqrt{h_{3,0}^{2}+h^{2}_{2,0}}}
\end{pmatrix}.\notag
\end{align}%
Therefore, interorbital E$_g$ pairings can be approximated by  the three-dimensional chiral $d$-wave state in the band basis.

\section{S5: Details of the employed BTK model}
The theoretical point-contact spectra were calculated by using the model for tunneling and Andreev reflection at the interface between a normal metal (N) and an anisotropic superconductor (S) \cite{Kashiwaya96,Yamashiro_PhysRevB.56}. The order parameter in the superconductor was assumed to have a simple dependence on $k_z$, of the form $\Delta(\mathbf{k}= \Delta_0 \sin \left(k_z/k\right)$ -- or, in spherical coordinates, $\Delta (\theta, \phi)=\Delta_0 \sin(\cos(\theta))$ where $\theta$ is the polar angle (the $z$ axis coinciding with the $k_z$ direction). Due to the cylindrical symmetry, there is no dependence on the azimuthal angle $\phi$. To account for the almost perfectly two-dimensional nature of the Fermi surface, we chose to adopt the same approach as in Ref.\cite{Yamashiro_PhysRevB.56}, i.e. we restricted the $z$ component of the Fermi surface to a cylindrical equatorial region defined by $\pi/2 -\delta \le \theta \le \pi/2 + \delta$ with $\delta = \pi/10$. The Fermi wave number and the effective mass were assumed to be identical in the normal and superconducting banks.

\begin{figure}
    \centering
    \includegraphics[width=1\linewidth]{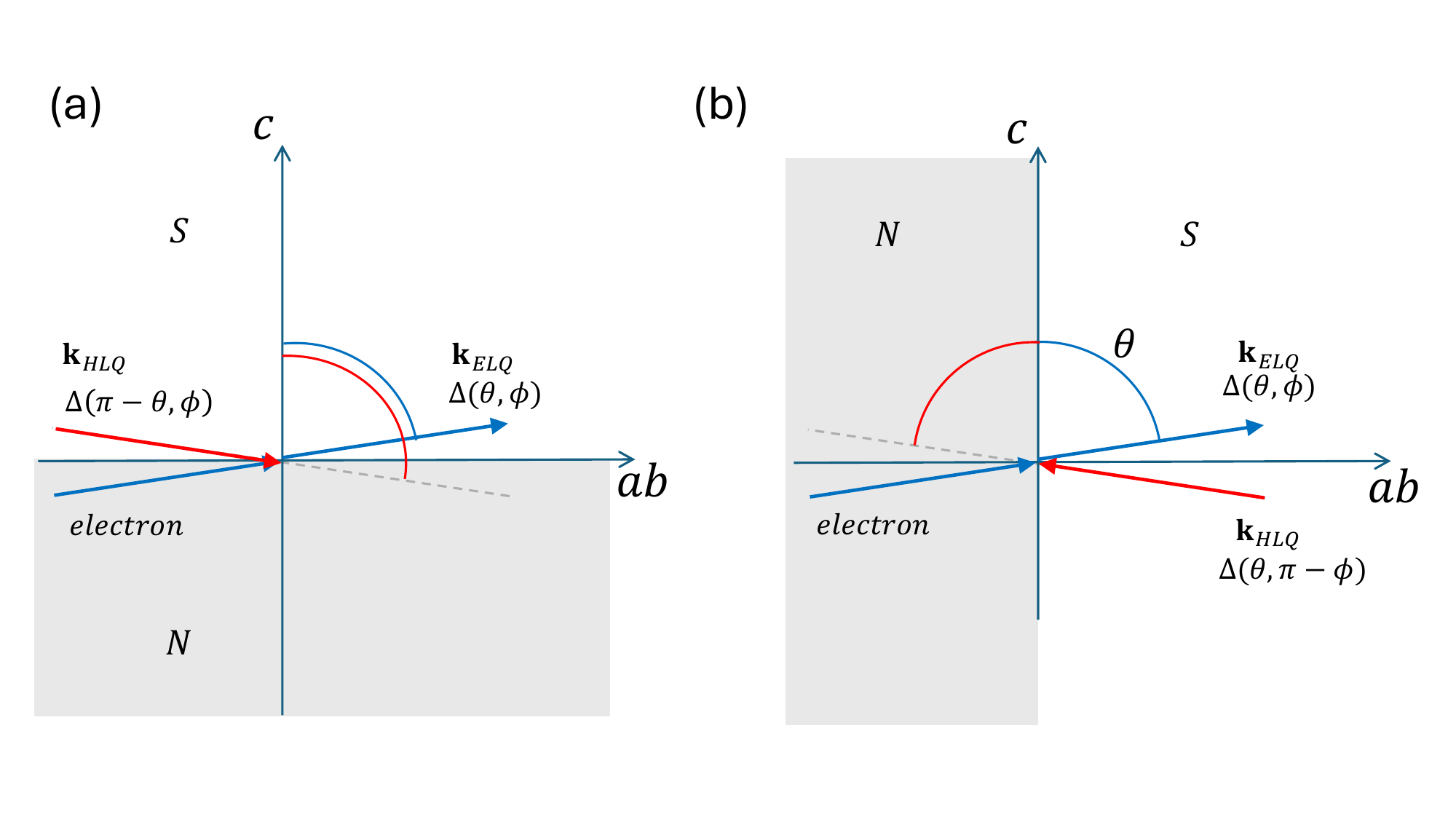}
    \caption{Scheme of the momenta of incoming electrons and transmitted quasiparticles (ELQ and HLQ) for $c$-axis contacts (panel a) and $a$-axis contacts (panel b). The label "$ab$" indicates the $ab$ plane. The relevant gap "seen" by the two kinds of quasiparticles is indicated. Due to the cylindrical symmetry about the $c$ axis, the change in $\phi$ (panel b) has no effect on the sign of the gap.}
    \label{fig:SuppFigBTK}
\end{figure}

For $c$-axis contacts (Figure \ref{fig:SuppFigBTK}a) the interface is represented by the $(x,y)$ plane. Electrons impinging at a certain angle $\theta$ (close to $\pi/2$) with respect to the $z$ axis will be either reflected as an electron, retro-reflected as a hole, transmitted as electronlike quasiparticle (ELQ) or transmitted as holelike quasiparticle (HLQ). ELQ and HLQ will have the same in-plane component of the momentum, and the same $k_z$ (for ELQ) or opposite $k_z$ (for HLQ) as the incoming electron. Consequently, HLQ and ELQ are sensitive to order parameters along different directions, i.e. $\Delta(\theta, \phi)$ for ELQ and $\Delta (\pi-\theta, \phi)$ for HLQ. These have the same absolute value but opposite sign, and the consequent $\pi$ phase shift is responsible for the formation of zero-energy bound states (ABS) at the interface \cite{KashiwayaPRB1995}. 

Instead, for $a$-axis contacts, the interface is represented by the $(y,z)$ plane and an electron impinging at an angle $\theta$ can be transmitted as a HLQ or ELQ that feel the order parameters $\Delta(\theta, \phi)$ and $\Delta(\theta, \pi-\phi)$ that are actually identical. This prevents the formation of ABS along this direction. 

The complete calculation requires an integration of the one-dimensional conductance in the region of the Fermi surface accessible to incoming electrons \cite{Daghero2011,Yamashiro_PhysRevB.56}. In particular, for $c$ axis contacts the normalized conductance is given by
\begin{equation}
\sigma_c(E)=\frac{\int_0^{2\pi} \int_{\pi/2}^{\pi/2+\delta} \sigma_{S,c} (E, \theta, \phi)\cos \theta \sin \theta \,d\theta \,d \phi}{2\pi \int_{\pi/2}^{\pi/2+\delta}\sigma_{N,c}(\theta) \cos \theta \sin \theta \, d \theta}
\end{equation}
where $\sigma_{S,c}(E, \theta, \phi)$ is the one-dimensional conductance calculated for the energy $E$ and along the direction defined by the angles $\theta$ and $\phi$ in reciprocal space
and $\sigma_{N,c}(\theta)$ is the corresponding normal-state conductance, which reads
\begin{equation}
    \sigma_{N,c}(\theta) =\frac{\cos^2 \theta}{\cos^2 \theta + Z^2}.
\end{equation}

In the case of $a$-axis contacts, the conductance is given by
\begin{equation}
\sigma_a(E)=\frac{\int_{\pi/2}^{\pi/2+\delta} \int_{-\pi/2}^{\pi/2} \sigma_{S,a} (E, \theta, \phi) \cos \phi \, d\phi \sin^2 \theta \, d\theta} {\int_{\pi/2}^{\pi/2+\delta}\int_{-\pi/2}^{\pi/2}\sigma_{N,a}(\theta, \phi) \cos \phi \, d\phi\,\sin^2\theta \, d \theta}
\end{equation}
where, again, $\sigma_{S,a}(E, \theta, \phi)$ is the one-dimensional conductance calculated for the energy $E$ and along the direction $(\theta,\phi)$ in reciprocal space \cite{Yamashiro_PhysRevB.56}, and $\sigma_{N,a}(\theta,\phi)$ is the corresponding normal-state conductance, i.e.
\begin{equation}
    \sigma_{N,a}(\theta, \phi)=\frac{\cos^2 \phi \, \sin^2 \theta }{\cos^2 \phi \, \sin^2 \theta + Z^2}.
\end{equation}

The one-dimensional conductances in the superconducting state, i.e. $\sigma_{S,c}$ and $\sigma_{S,a}$, contain information about the order parameter along the momenta of hole-like quasiparticles and electron-like quasiparticles, and the phase difference between the two, as already pointed out before and as described in detail in ref. \cite{Yamashiro_PhysRevB.56}.

\bibliography{biblio_SM}